# Tools for Multiaxial Validation of Behavior Laws Chosen for Modeling Hyper-elasticity of Rubber-like Materials

L. Chevalier* and Y. Marco

LMT-Cachan
ENS de Cachan / CNRS – UMR 8535 / Université Paris 6
61 avenue du Président Wilson
94235 Cachan cedex
France

* to whom correspondence should be addressed
Tel: +33 1 47 40 21 76
Fax: +33 1 47 40 22 40
E-mail : chevalier@lmt.ens-cachan.fr

# Tools for Multiaxial Validation of Behavior Laws Chosen for Modeling Hyper-elasticity of Rubber-like Materials

By:

L. Chevalier* and Y. Marco

***Abstract***

We present an experimental approach to discriminate hyper-elastic models describing the mechanical behavior of rubber-like materials. An evaluation of the displacement field obtained by digital image correlation allows us to evaluate the heterogeneous strain field observed during these tests. We focus on the particular case of hyper-elastic models to simulate the behavior of some rubber-like materials. Assuming incompressibility of the material, the hyper-elastic potential is determined from tension and compression tests. A biaxial loading condition is obtained in a multiaxial testing machine and model predictions are compared with experimental results.

# 1. Introduction

The variety of constitutive models representing the hyper-elastic behavior of rubber-like materials, it is quite difficult to decide which one represents the material responses of the material to different loads most accurately. Since Mooney (1), Rivlin (2) and Treolar (3), many publications (see Arruda and Boyce (4) or Heuillet et al (5)) have proposed different approaches that we can divide into two classes, as the models “based upon physical consideration” and the “phenomenological” models. In this paper we will focus on the second kind of models.

Furthermore, experimental data have been produced to test the validity of these models. Using essentially two specific tools developed by LMT-Cachan, our aim is to make a critical review of classical tests and to propose a simple approach to prove the accuracy of each model. The first tool is the auto-correlation technique used to analyze displacement field during the loading of specimen. This technique allows us to quantify the strain field homogeneity for any kind of loading. The second and complementary tool is a multiaxial testing machine named ASTREE (manufactured by Schenck) which allows multiaxial loading on 2D or 3D specific specimens.

In the first part, we will illustrate the combined use of these tools to analyze and validate homogeneous tension and compression tests. We present the hyper-elastic potential identification on these two tests and we discuss the accuracy of some classical forms. In the second part, we validate the identification made by comparing numerical simulations made using the ABAQUS code with the experimental data produced by the biaxial loading tests.

## 1.1 A rubber like material

The rubber used for the tests was Smactane$^{TM}$ which is produced by SMAC (Toulon – France). This rubber exhibits a maximum damping over a large range of temperature (i.e., from -50° to +120°c). reversible mechanical behavior to over 800% elongation and a breaking strength of 9.5 MPa (Piola-Kirchhof stress). In the following tests, we will limit levels to ensure that no other effect than hyper-elasticity (such as irreversible effects, damage) occurs.

## 1.2 A hyper-elastic potential approach to model rubber-like material

The macroscopic approach of homogeneous, hyper-elastic materials such as rubber-like materials proposed by Rivlin (2) consists in the introduction of an elastic potential. Classical hypotheses of isotropy, material-frame indifference and incompressibility allows one to assume that the state potential W only depends on the first two invariant $I_1$ and $I_2$ of the right Cauchy-Green tensor $\underline{\underline{C}}$ (where $\underline{\underline{C}} = {}^T\underline{\underline{F}}.\underline{\underline{F}}$, $\underline{\underline{F}}$ being the transformation gradient):

$$\left.\begin{aligned} I_1 &= \mathrm{tr}\underline{\underline{C}} = \lambda_1^{\,2} + \lambda_2^{\,2} + \lambda_3^{\,2} \\ I_2 &= \frac{1}{2}\left(\mathrm{tr}^2\underline{\underline{C}} - \mathrm{tr}\underline{\underline{C}}^2\right) = \lambda_1^{\,2}\lambda_2^{\,2} + \lambda_2^{\,2}\lambda_3^{\,2} + \lambda_3^{\,2}\lambda_1^{\,2} \end{aligned}\right\} \text{ with } I_3 = \det\underline{\underline{C}} = \lambda_1\lambda_2\lambda_3 = 1 \qquad (1)$$

where $\lambda_1$, $\lambda_2$, $\lambda_3$ are the three principal extension ratios. With such a potential and the use of conventional formalism of continuum thermodynamics, the hyper-elastic constitutive law derives from W:

$$\underline{\underline{S}} = \frac{\partial W}{\partial \underline{\underline{E}}} = 2\frac{\partial W}{\partial \underline{\underline{C}}} = 2\left( \underline{\underline{1}}\frac{\partial W}{\partial I_1} - \underline{\underline{C}}^{-2}\frac{\partial W}{\partial I_2} \right) \quad (2)$$

where $\underline{\underline{E}}$ is Green Lagrange strain and $\underline{\underline{S}}$ the second Piola Kirchhoff extra stress tensor. The Cauchy extra stress tensor $\underline{\underline{\Sigma}}$ is related to the Piola Kirchhoff extra stress tensor by:

$$\underline{\underline{\Sigma}} = \underline{\underline{F}}.\underline{\underline{S}}.{}^t\underline{\underline{F}} = 2\left( \underline{\underline{B}}\frac{\partial W}{\partial I_1} - \underline{\underline{B}}^{-1}\frac{\partial W}{\partial I_2} \right) \quad (3)$$

We obtain the complete Cauchy stress tensor $\underline{\underline{\sigma}}$ by the relation:

$$\underline{\underline{\sigma}} = \underline{\underline{\Sigma}} - p\underline{\underline{I}} \quad (4)$$

where p is the pressure associated with the incompressibility condition. Since the partial derivatives of W with respect to $I_1$ and $I_2$ are known, the behavior is completely defined when it remains hyper-elastic.

### *1.3 Tests to manage accurate form of the hyper-elastic potential*

The different hyper-elastic potential forms proposed by Mooney (1), Mark (6), Gent and Thomas (7), Hart-Smith (8), Alexander (9) and even more recently by Lambert-Diani and Rey (10) for example, lead to uncoupled expressions of the potential W partial derivatives with respect to $I_1$ and $I_2$:

$$\frac{\partial W}{\partial I_1} = f(I_1) \quad \text{and} \quad \frac{\partial W}{\partial I_2} = g(I_2) \quad (5)$$

Other authors (McKenna et al. (11) for example) directly found these derivatives from measurement of torque and normal force during a torsion test. Our approach considers homogeneous tests: uniaxial tension and compression, to identify these derivatives. Hyper-elastic models give the following Cauchy stress $\sigma$ in uniaxial tension and compression wrt. Longitudinal elongation $\lambda$ :

$$\sigma = 2f(I_1)\left(\lambda^2 - \frac{1}{\lambda}\right) - 2g(I_2)\left(\frac{1}{\lambda^2} - \lambda\right) \quad \text{with} \begin{cases} I_1 = \lambda^2 + \dfrac{2}{\lambda} \\ I_2 = \dfrac{1}{\lambda^2} + 2\lambda \end{cases} \quad (6)$$

Figure 1 shows that $I_1$ is greater than $I_2$ in uniaxial tension and that $I_2$ is greater than $I_1$ in compression. We can also see that pure shear or plane strain tests lead to identical values of $I_1$ and $I_2$ and that equi-biaxial strain leads to the same values of $I_1$ and $I_2$ as compression does. Since $I_1$ is greater than $I_2$ in uniaxial tension and that it is observed that f has higher values than g, an approximate identification of the function $f(I_1)$ from tension data is possible. The fitted function f can be used to determine function $g(I_2)$ from compression data. This is the method we will use to make the first approximation of the hyper-elastic potential in the following section.

Besides this class of potential form, Ogden (12) proposes a hyper-elastic potential expression depending of $\underline{\underline{C}}$ eigenvalues (i.e., square of principal elongation $\lambda_i^2$). The procedure proposed here is no longer suitable for identification. As Ogden model is often used in finite element codes (Abaqus for example) we will manage identification and numerical simulations for this model too.

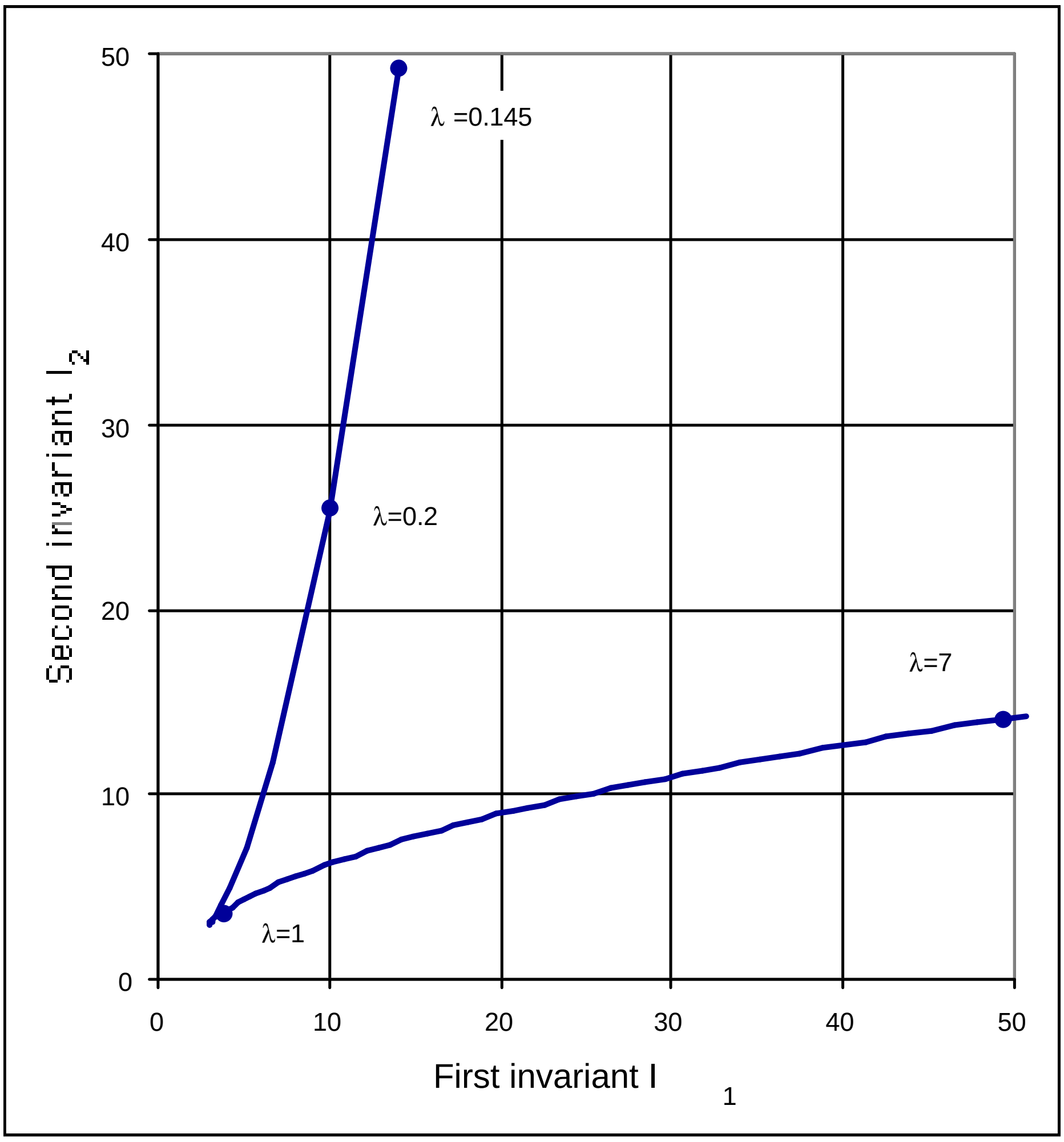

Figure 1: Position of invariant $I_1$ and $I_2$ during tension and compression tests.

## *2. Homogeneous tests in finite strain*

In this section we will introduce the strain field measurement method. We use this method to validate strain homogeneity during tension tests, and we explain how image analysis is used to test friction condition during compression tests. Following a method inspired from the identification procedure of Lambert-Diani et al. (10), both tests are used to identify the hyper-elastic potential expression.

### *2.1 Uniaxial tension test*

Tension tests are carried out on a Deltalab testing machine. The initial specimen dimensions are 80x30x5 $mm^3$. Both edges of the specimen are blocked in hydraulic clamps. The large amplitude of the traverse displacement makes it possible to reach large elongation up to 7. A CCD camera is used to take pictures of the specimen being deformed. After treatment of the series of pictures, we can plot the Cauchy tension stress $\sigma$ versus the axial elongation $\lambda_1$ where '1' is the tension direction).

#### **Strain field measurement: the cross correlation technique**

To determine the displacement field of one image with respect to a reference image, one considers a sub-image (i.e., a square region) which will be referred to as a “zone of interest” (ZOI). The aim of the correlation method is to match the zones of interest in both images (Fig. 2). The displacement of one ZOI with respect to the other one is a two-dimensional shift of an intensity signal digitized by a CCD camera. To estimate a shift between two signals, one of the standard approaches is to use a correlation function.

The theoretical aspects of the correlation are developed in a previous paper from Chevalier et al. (12). Two images are examined. The first one is referred to as ‘reference image’ and the second one is called ‘deformed image.’ The largest value p of a ‘region of interest’ (ROI) of size $2^p$ x $2^p$ centered in the reference image is selected. The same ROI is examined in the deformed image. A first FFT correlation is performed to determine the average displacement $U_0$, $V_0$ of the deformed image with respect to the reference image. This displacement is expressed in an integer number of pixels and corresponds to the maximum of the cross-correlation function evaluated for each pixel of the ROI. This first prediction enables one to determine the maximum number of pixels that belong to both images. The ROI in the deformed image is now centered on a point corresponding to displaced center of the ROI in the reference image by an amount $U_0$, $V_0$.

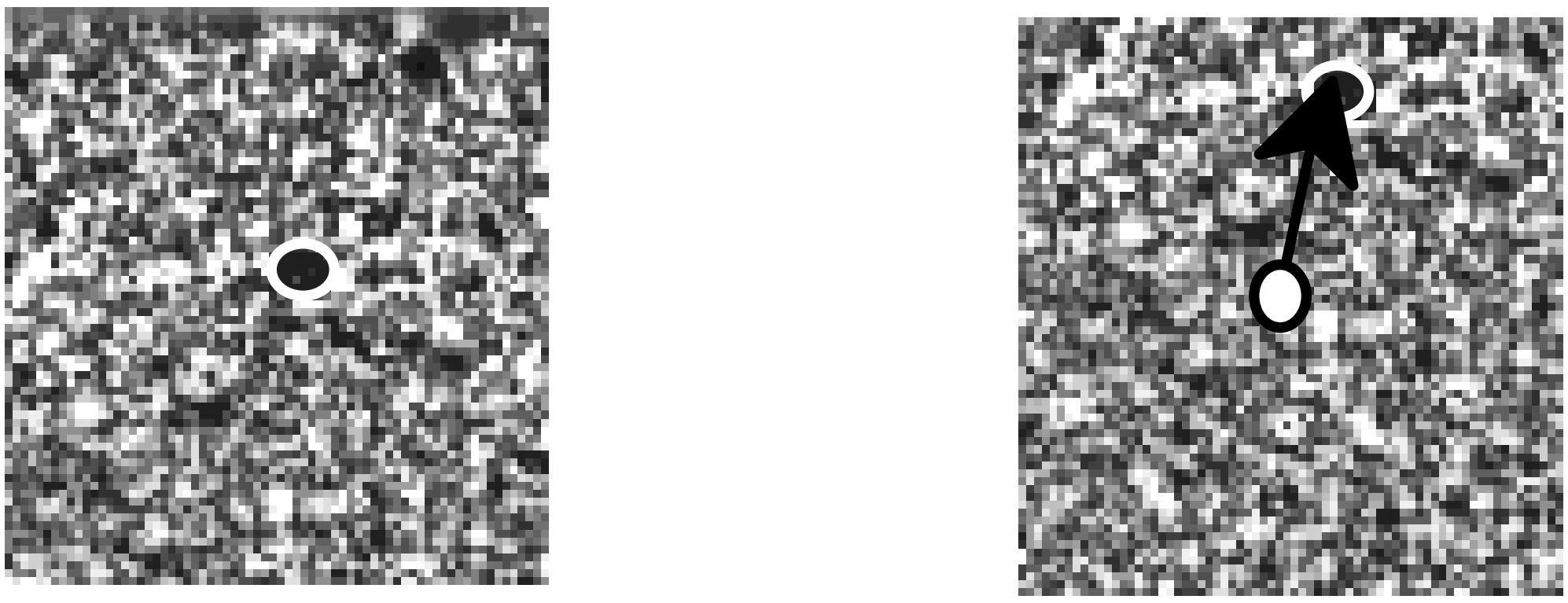

Figure 2: ZOI in the initial and ‘deformed’ image. The middle black spot of the left picture is located in the top area of the right picture. The vector represents the displacement of the ZOI center

The user usually chooses the size of the zones of interest (ZOI) by setting the value of $s < p$ so that the size is $2^s$ x $2^s$. To map the whole image, the second parameter to choose is the shift $\delta x$ (= $\delta y$) between two consecutive ZOI: $1 \leq \delta x \leq 2^s$. These two parameters define the mesh formed by the centers of each ZOI used to analyze the displacement field. The following analysis is performed for each ZOI independently. A first FFT correlation is carried out and a first value of the in-plane displacement correction $\Delta U$, $\Delta V$ is obtained. The values $\Delta U$, $\Delta V$ are once more integer numbers so that the ZOI in the deformed image can be displaced by an additional amount $\Delta U$, $\Delta V$. The displacement residual s are now less than 1/2 pixel in each direction. A sub-pixel iterative scheme can be used. The procedure, CORRELI$^{GD}$ proposed by Hild et al. (14) is implemented in Matlab$^{TM}$. The precision of the method is at least on the order of 2/100 pixel at least and the minimum detectable displacement is also on the order of 1/100 pixel. A typical result is shown in Fig. 3 where the vertical displacement component contours are plotted on superposition to the reference image.

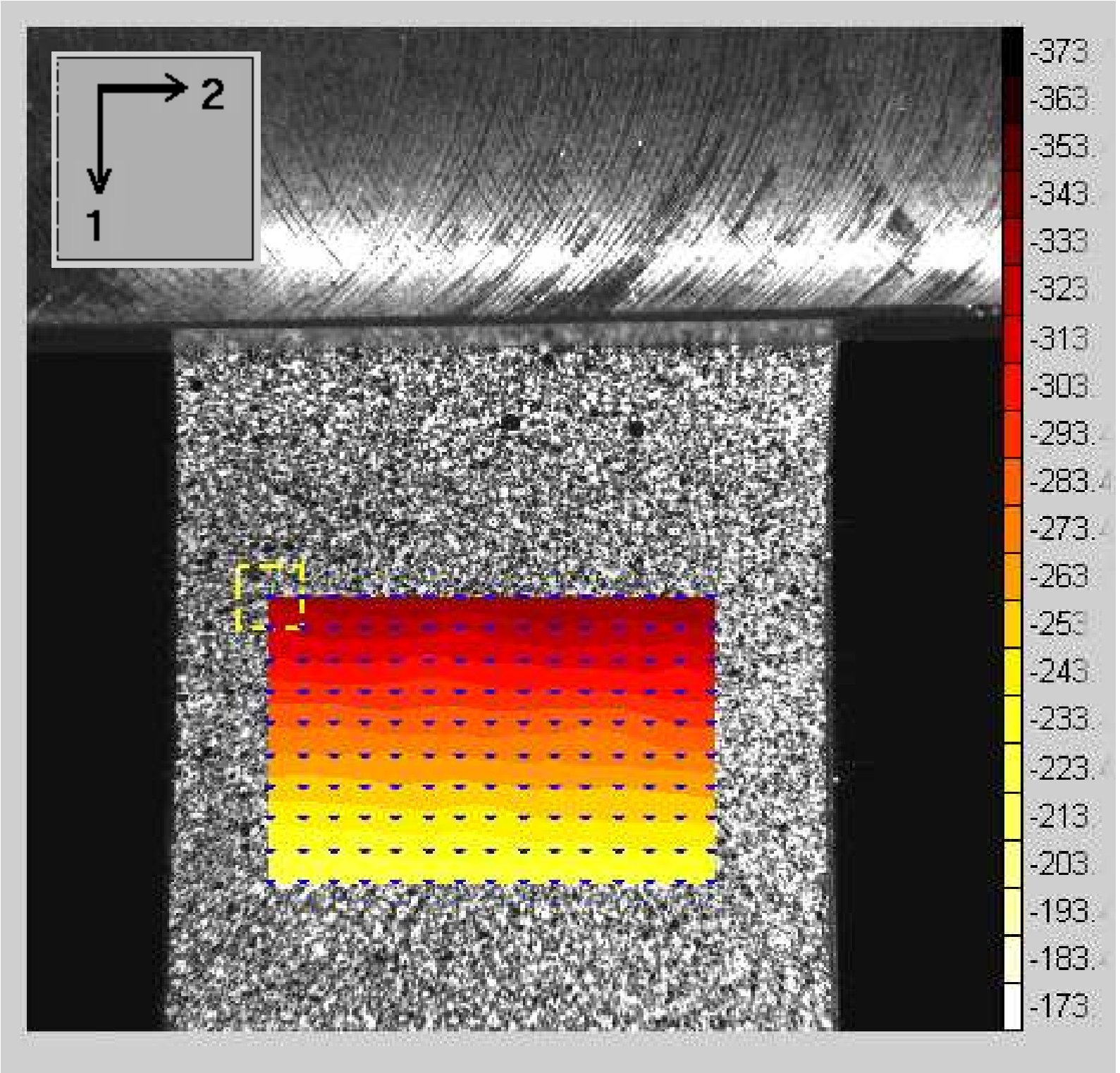

Figure 3: Typical result for uniaxial tension. $U_1$ displacement contours (in pixels) are regularly spaced which is characteristic of a homogeneous strain field.

The problem now is that the initial grid can move out of the picture so that for very large strains, we must initially choose a suitable mesh and then evaluate its displacement step by step to produce the tension curve, or move the CCD camera according to the previous displacement value.

#### Tension behavior: incompressibility and specimen sliding

Such a technique is very helpful, especially in the case of large deformation for which no strain gauge can keep up with the elongated specimen. Furthermore, the displacement measured on the machine traverse includes a sliding effect in the grips and cannot allow the calculation of true strain. As we can see Fig. 4-b, the strain calculated using machine displacement gives accurate results for the first 20 % of deformation. After that, progressive sliding occurs in the clamps and the real strain of the material is lower than if it were estimated with a classical technique.

Another useful piece of information given by the image analysis is the transversal strain that is also evaluated during the test. On Fig. 4-a, the equivalent of the Poisson ratio $\upsilon$ (defined arbitrarily as the ratio of longitudinal nominal strain with the transversal nominal strain) is plotted in continuous line, assuming isotropy ($\lambda_2 = \lambda_3$) and incompressibility ($\lambda_1.\lambda_2.\lambda_3 = 1$). The expression of this ratio is given by:

$$\upsilon = -\frac{\lambda_2 - 1}{\lambda_1 - 1} = \frac{1 - 1/\sqrt{\lambda}}{\lambda - 1} \quad (7)$$

$\lambda_1$ will be named $\lambda$ in the following paragraphs. The circles represent experimental values obtained during the tension test. Since both curves almost fit, we can consider that both assumptions were true. If it was not the case, a simple apparatus could help to make tell if the material is compressive or presents anisotropy by measuring $\lambda_3$. Two ways can be tested: using a single CCD camera with a viewpoint in 45° direction wrt. The orientation of the principal face ($X_1$, $X_2$) of the specimen; using two CCD camera viewing both $X_1$, $X_2$ and $X_1$, $X_3$ faces. In that case the two cameras must be synchronized. This second direction is actually tested.

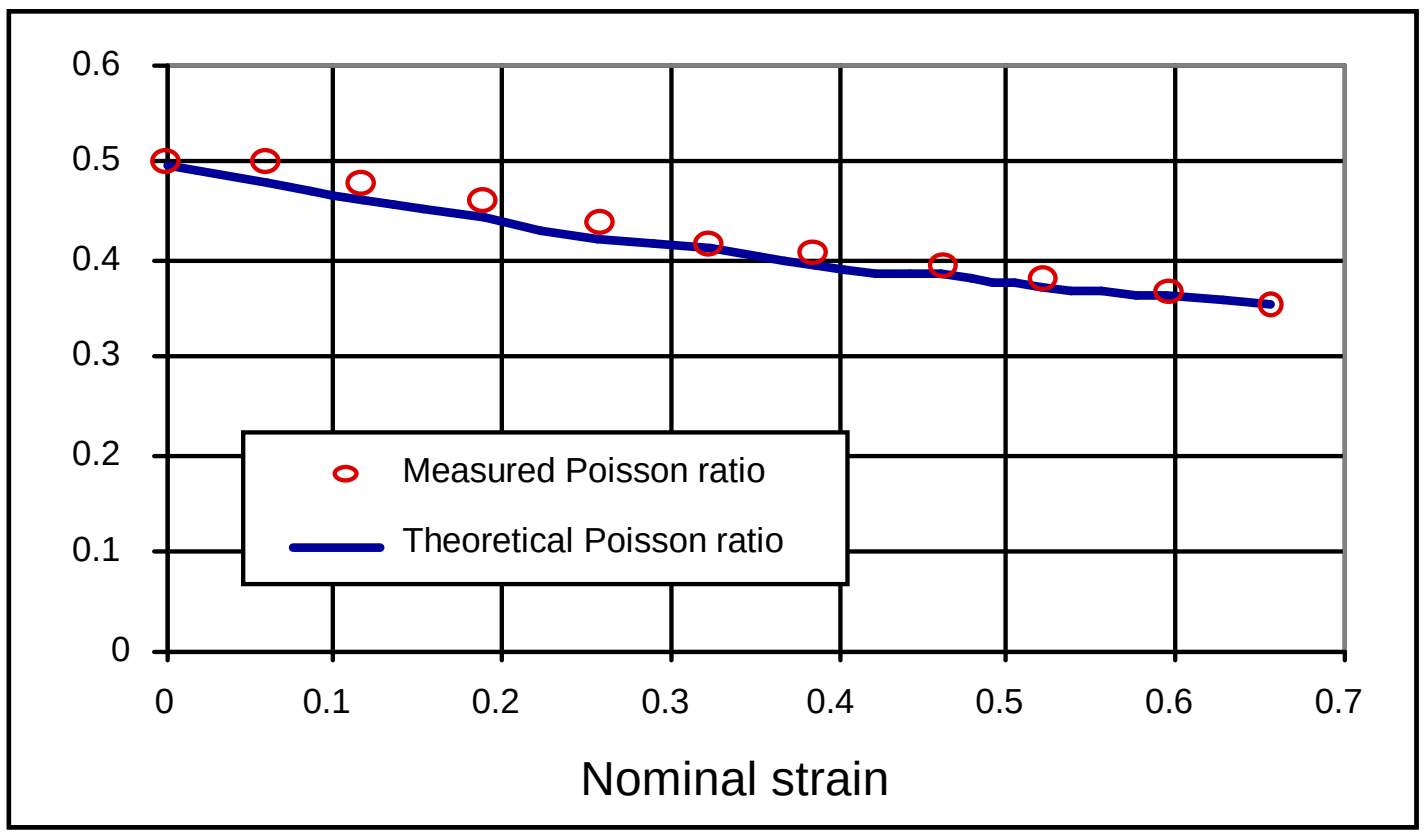

(a)

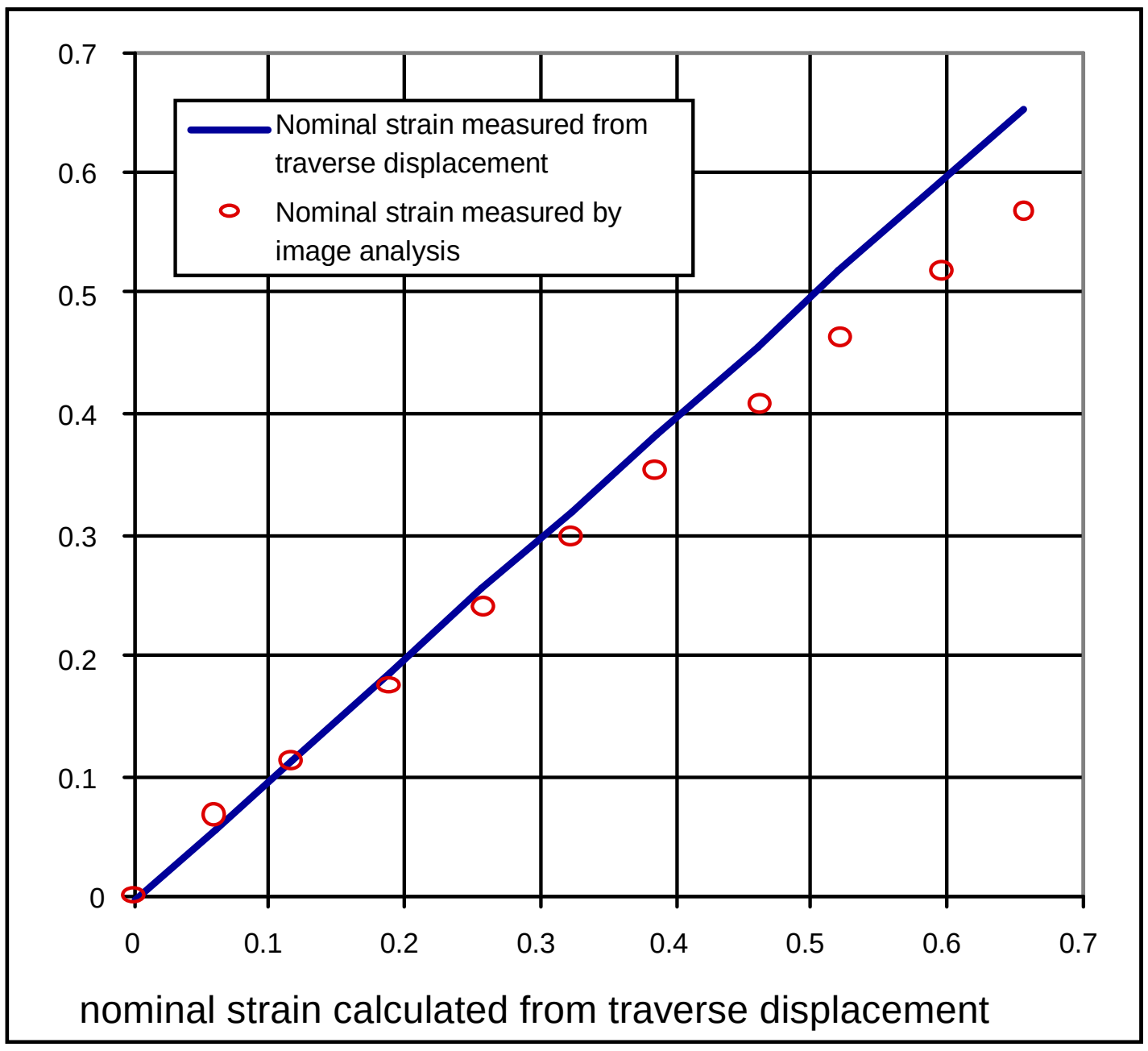

(b)

Figure 4: Strain measured from image analysis. (a) the Poisson effect confirms the incompressibility assumption, (b) global strain (i.e., measured from traverse displacement) and true strain (i.e., measured by image analysis) show regular sliding in grip since 20% of deformation.

Finally, the typical displacement field shown on Fig. 3 shows that both longitudinal and transversal components are linear versus the coordinates $x_1$. Consequently, the homogeneity of the strain field is verified for this tension test. Finally, experimental data obtained on our rubber specimen loaded in uniaxial tension are shown in Fig. 5. Elongation $\lambda$ is calculated from the strain measured by image analysis. The Cauchy tension stress $\sigma$ is calculated from the tension force T reduced to the current section S.

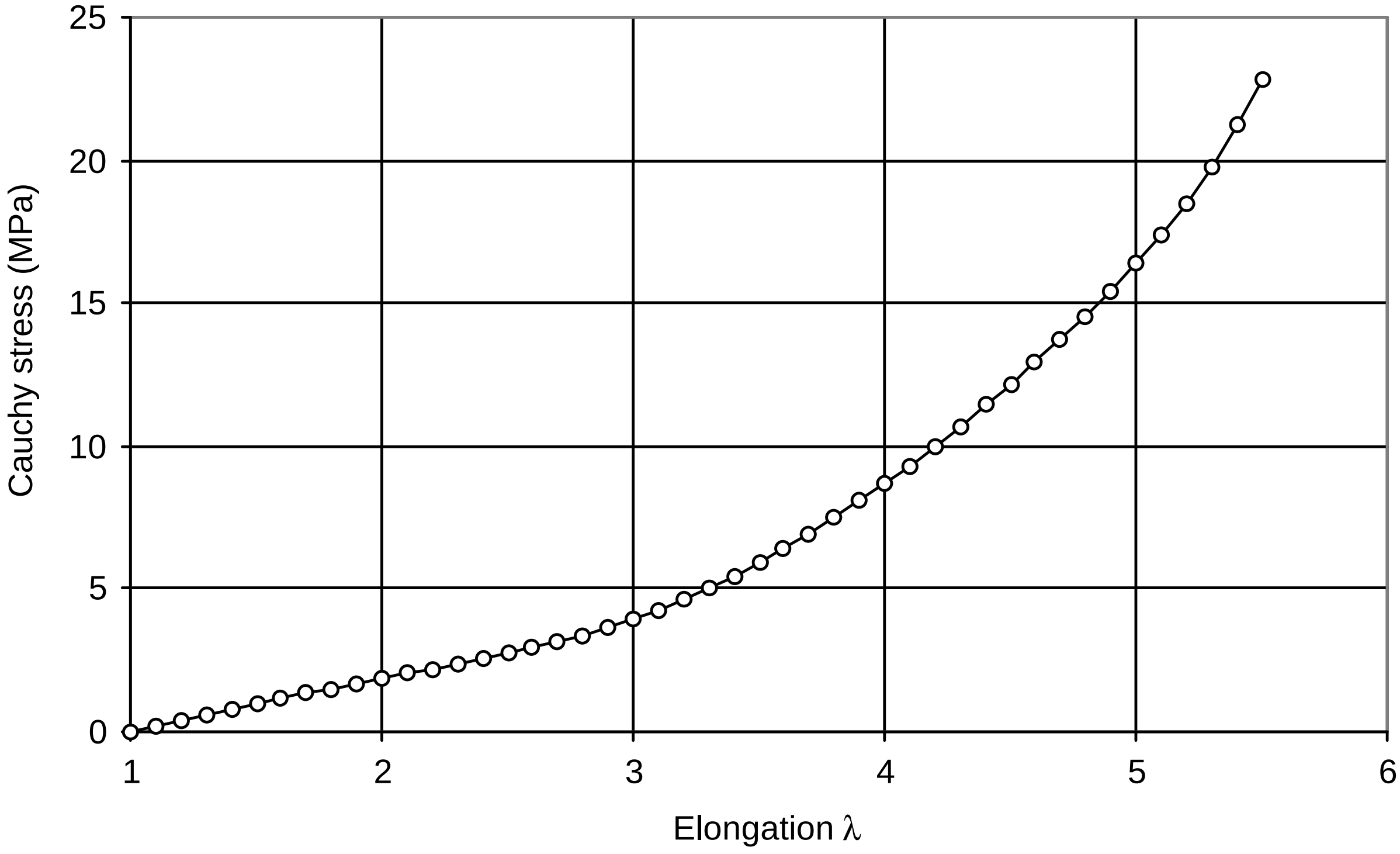

Figure 5: Stress/strain response of Smactane$^{TM}$ in uniaxial tension.

## *2.2 Uniform compression test*

In order to identify the influence of the second invariant $I_2$, we carried out compression tests on small (10x10x5 $mm^3$) specimens. The compression apparatus is fixed on a MTS uniaxial machine. The lower compression plate is made of glass so that it is possible to follow the expansion in both perpendicular directions. The 45° mirror fixed below this glass compression plate make it easier to record the expansion evolution. The upper plate is fixed to the load cell that measures the compression force.

### Compression apparatus: sticking or sliding control of the specimen

The main difficulty is to keep a homogeneous strain field in the specimen during the compression test. Figure 6-a shows the strain components $\varepsilon_{22}$ and $\varepsilon_{33}$ for a compressive strain $\varepsilon_{11}$ equal to 20%. Both values are less or equal to 1%, which clearly indicates that the specimen sticks on the compression plates. In figure 6-b, a low viscosity lubricant between the plates and the specimen leads to radial expansion. For a 20% compression both values of $\varepsilon_{22}$ and $\varepsilon_{33}$ are equal to approximately 10%, which is compatible with the incompressibility assumption. In this second case, we can be sure that compression is homogeneous in the specimen.

**Compression behavior**

The Cauchy stress is plotted versus elongation in Fig. 7 (circles) for a range of elongation varying from 0.3 to 1. We verify that the same slope can be measured from tension and compression testing for elongation equal to 1. We also verify the hardening effect that occurs when $\lambda$ becomes very small. As a comparison we plotted in continuous line the prediction of compression stress issued from the potential identification from tension data only (i.e., as if the contribution of $g(I_2)$ was neglected). It confirms that the effect of the second invariant has to be taken into account. This leads to more complex form of the potential that makes difficult the identification procedure.

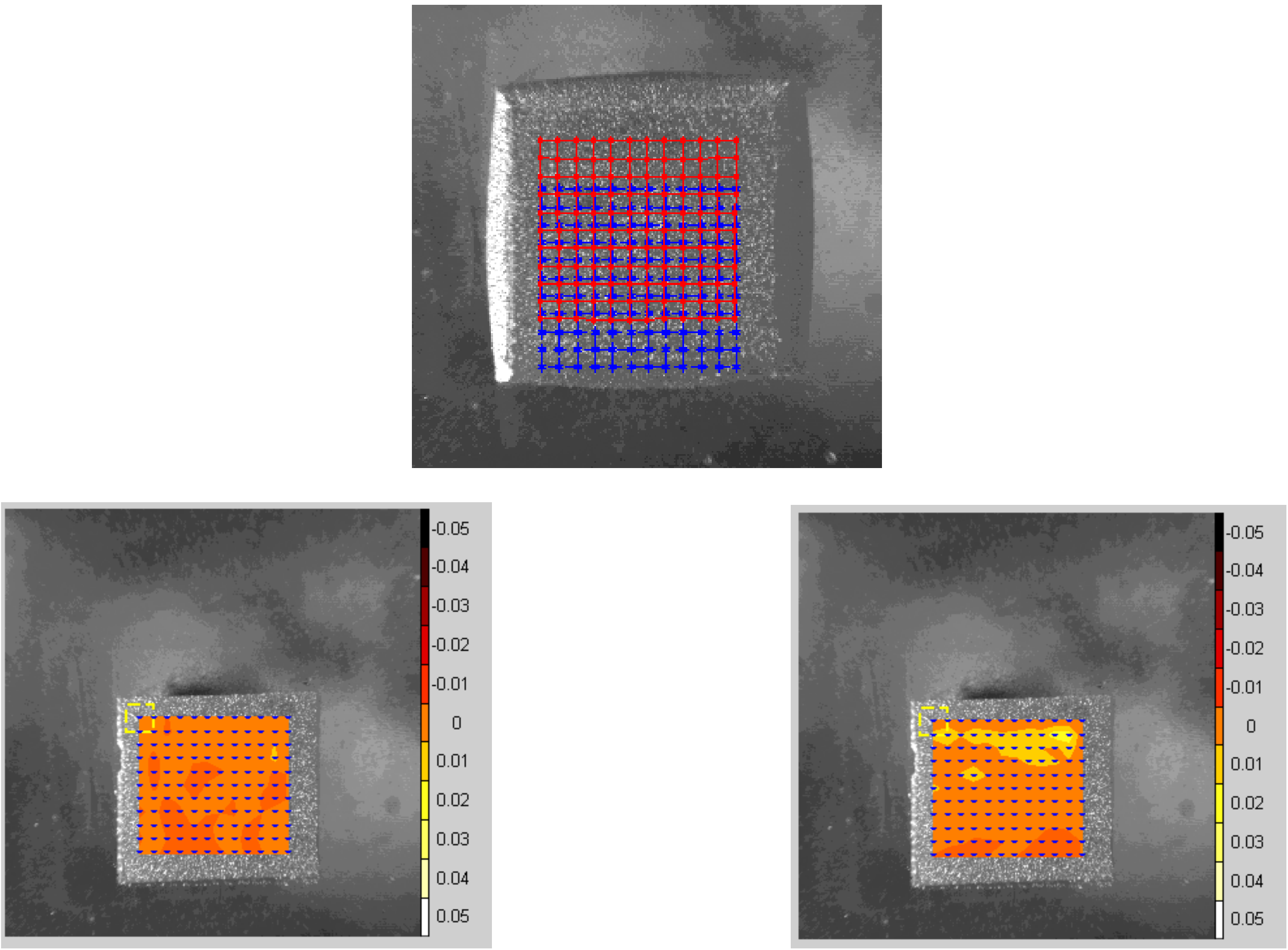

(a) Initial and displaced grid: the translation of the mirror is well reproduce between the two images. Both transversal strains are less than 1%

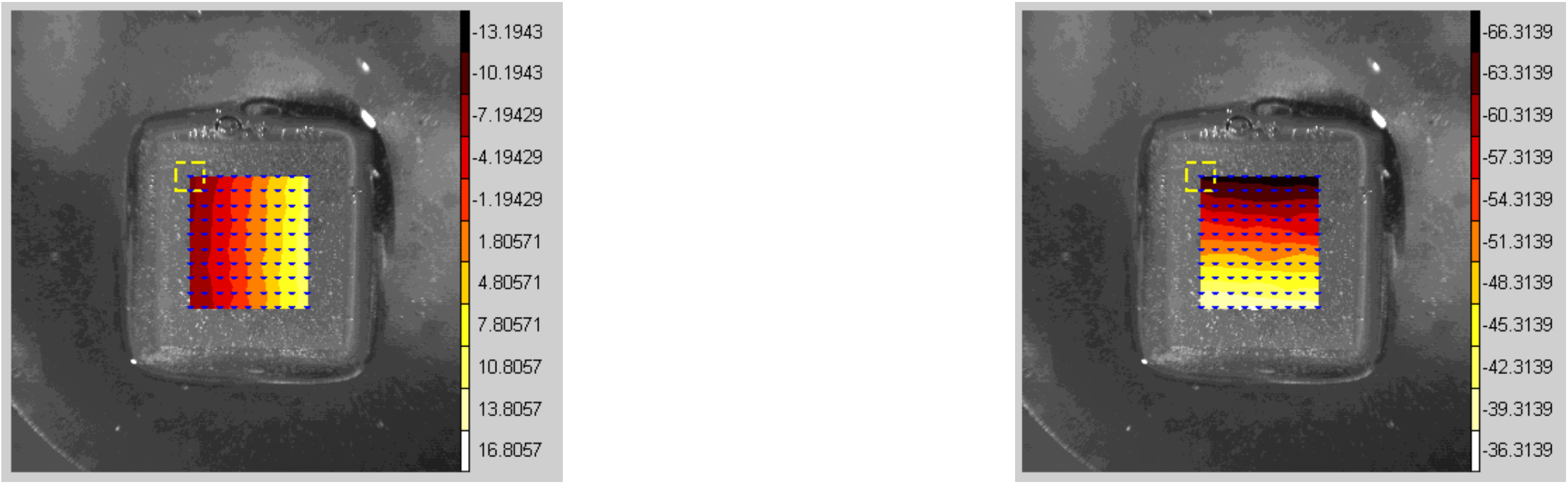

(b) Displacement components $U_2$ (left) and $U_1$ (right).

Both components lead to stain about 9% in both directions

Figure 6: Image analysis results in compression.

(a) compression with no lubricant → specimen sticking

(b) compression with lubricant → specimen sliding

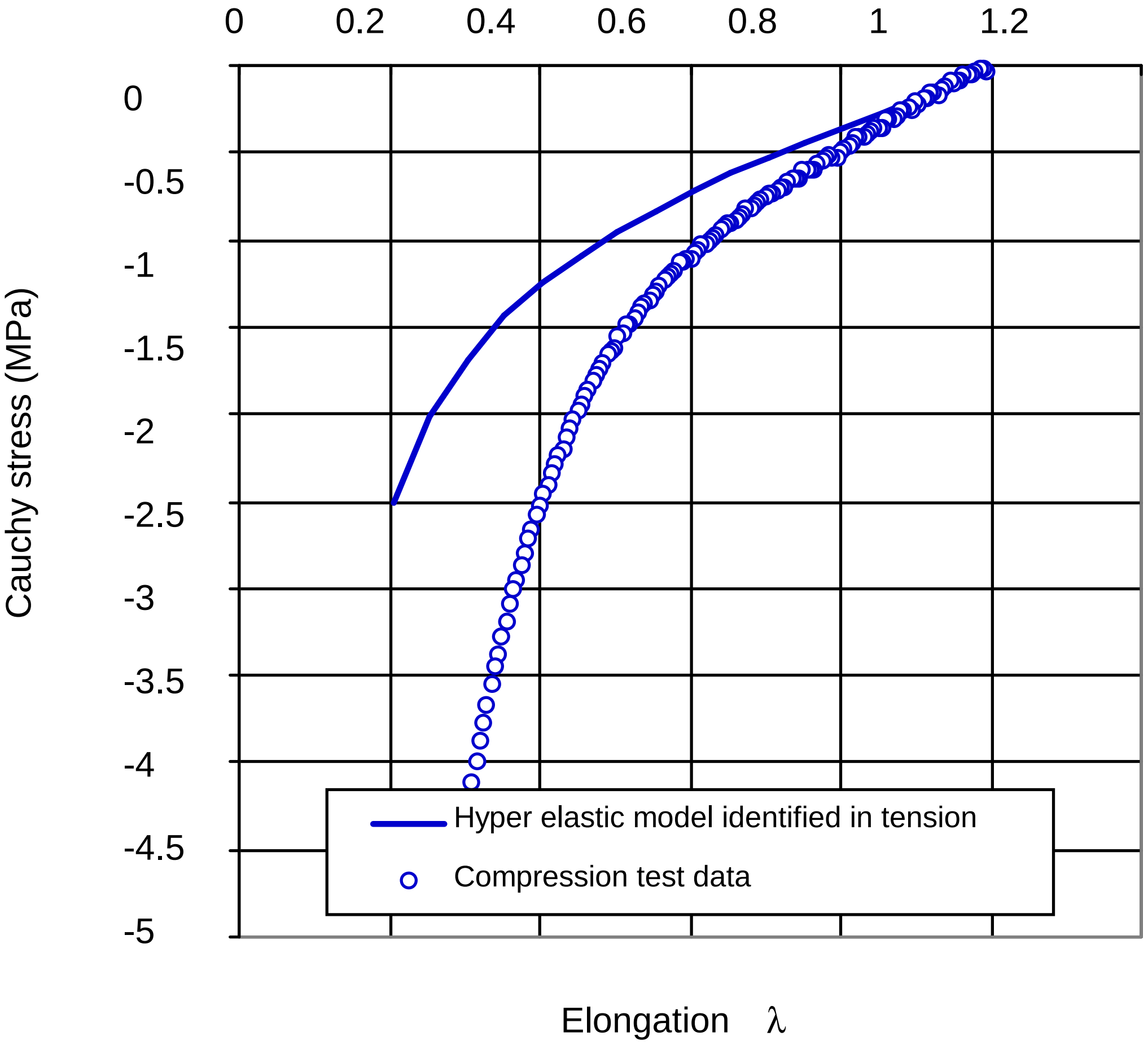

Figure 7: Compression test data (circles) and prediction with a potential identified in uniaxial tension only (continuous line)

### *2.3 Identification procedure*

In this section, we present the hyper-elastic potential procedure. As expressions can be complex, the identification by a simple minimization of the distance Δ between analytical expression and experimental data is not unique. The distance expression is given by:

$$\Delta = \frac{\sum_{i=1}^{N} \left(\sigma(\lambda_i) - \sigma_i\right)^2}{\sum_{i=1}^{N} \sigma_i^{\,2}} \qquad (8)$$

$\sigma_i$ is the measured stress for elongation $\lambda_i$ and $\sigma(\lambda_i)$ is the calculated stress for the same elongation using a given analytical form. This expression may have a single minimum for the analytical form proposed by Mooney, but it is not the case for more complex forms such as that of Gent and Thomas or that of Hart-Smith. It is therefore important to have a good approximation of functions $f(I_1)$ and $g(I_2)$ to begin the minimization process.

#### First invariant function identification

If we assume that the contribution of the second invariant is negligible with respect to the first one on the uniaxial tension stress $\sigma_T$ (since $I_1 >> I_2$), the function f can be directly identified from the knowledge of the Cauchy stress and extension ratio $\lambda$ (see Eqn. 6).

$$f(I_1) = \frac{\sigma_T}{2\left(\lambda^2 - \frac{1}{\lambda}\right)} \quad \text{with} \quad I_1 = \lambda^2 + \frac{2}{\lambda} \quad \text{and} \quad \lambda > 1 \tag{9}$$

Figure 8 shows that this function is not constant during the test. On Fig.8-a we try to fit with polynomial functions (i.e. Rivlin form assuming uncoupled evolution of f and g functions):

$$f(I_1) = \sum_{i=0}^{n} a_i (I_1 - 3)^i \tag{10}$$

On Fig.8-b we look for best fit with an exponential form:

$$f(I_1) = \exp\left\{\sum_{i=0}^{n} a'_i (I_1 - 3)^i\right\} \tag{11}$$

$a_i$ or $a'_i$ are material dependent parameters. We choose to keep the value n=2 for the first form because the experimental data are well described and the number of parameters is minimal, even though the value n=3 would fit the data better. This choice leads to the compression prediction plotted on Fig. 7 which is non representative. The second invariant function (i.e., function g) is clearly needed in order to fit experimental data as well in compression as in tension.

#### Second invariant function identification

With the first invariant function f estimated, we are able to plot the second invariant function g from the compression stress values $\sigma_C$ from Eqn.12.

$$g(I_2) = \frac{\sigma_C - 2f(I_1)\left(\lambda^2 - \frac{1}{\lambda}\right)}{2\left(\lambda - \frac{1}{\lambda^2}\right)} \quad \text{with} \quad I_1 = \lambda^2 + \frac{2}{\lambda}; I_2 = 2\lambda + \frac{1}{\lambda^2} \quad \text{and} \quad 0 < \lambda < 1 \tag{12}$$

Figure 9 shows the second function versus the inverse of $I_2$ (Fig.9-a) or versus the shifted second invariant (Fig.9-b). This function is, once again, not constant. In both cases the best fit leads to the following expressions:

$$g(I_2) = \sum_{i=0}^{n} \frac{b_i}{I_2^i} \tag{13}$$

$$g(I_2) = \sum_{i=0}^{n} b'_i (I_2 - 3)^i \tag{14}$$

$b_i$ or $b'_i$ are material dependent parameters. We choose to keep the expression of Eqn. 13 which is analogous of the Hart-Smith form with n=2.

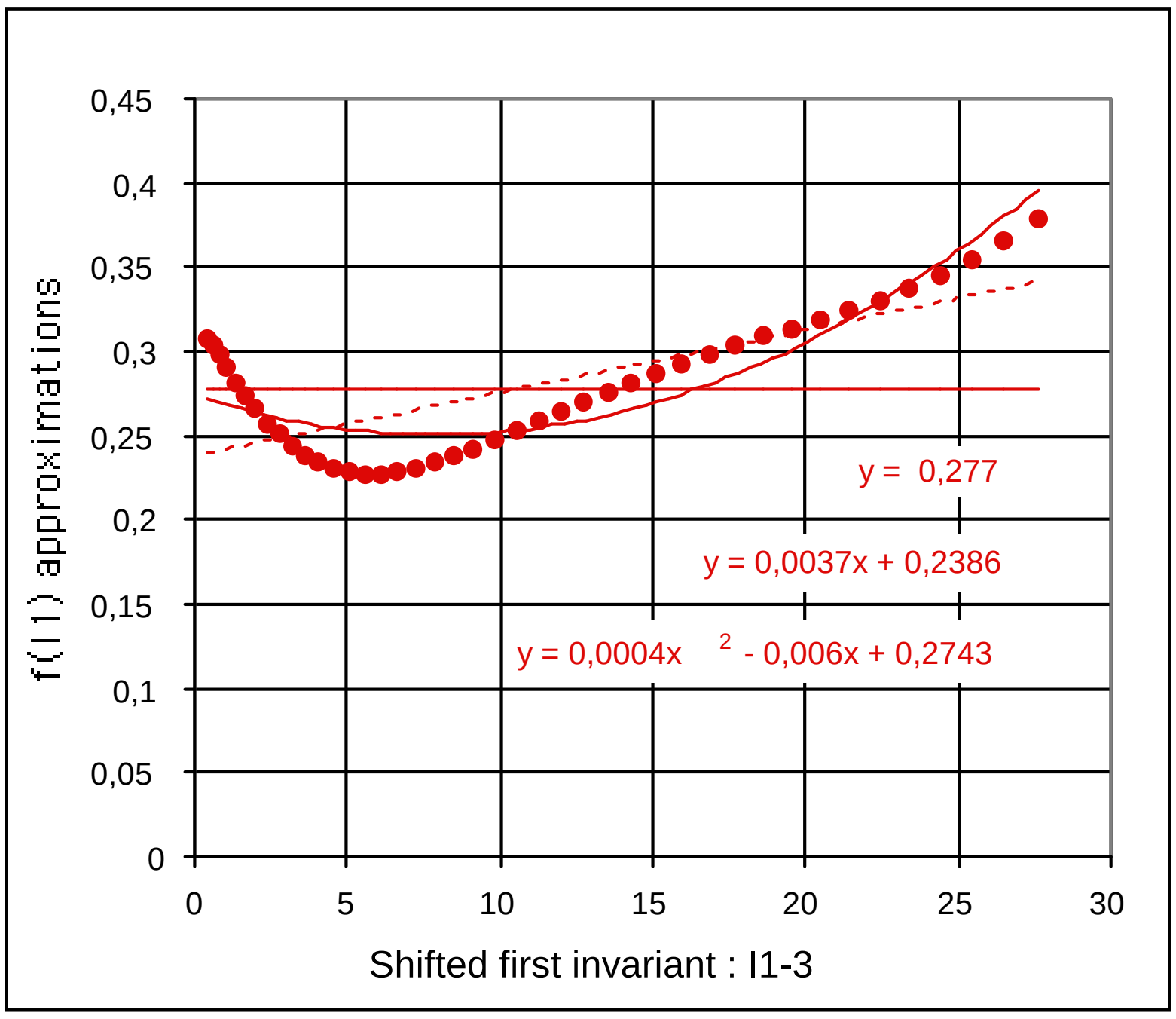

(a)

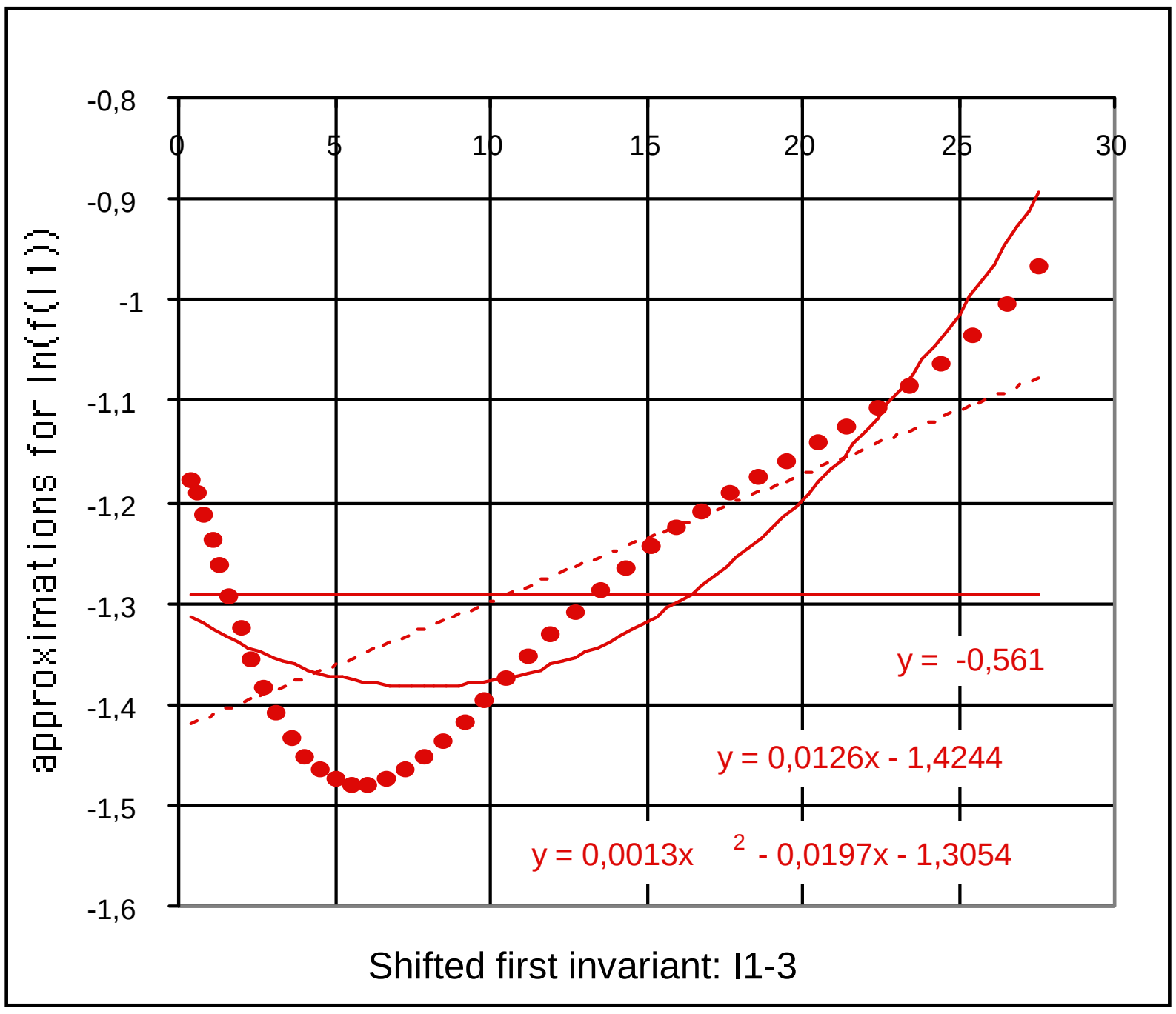

(b)

Figure 8: First invariant identification.

(a) Rivlin form,

(b) exponential form.

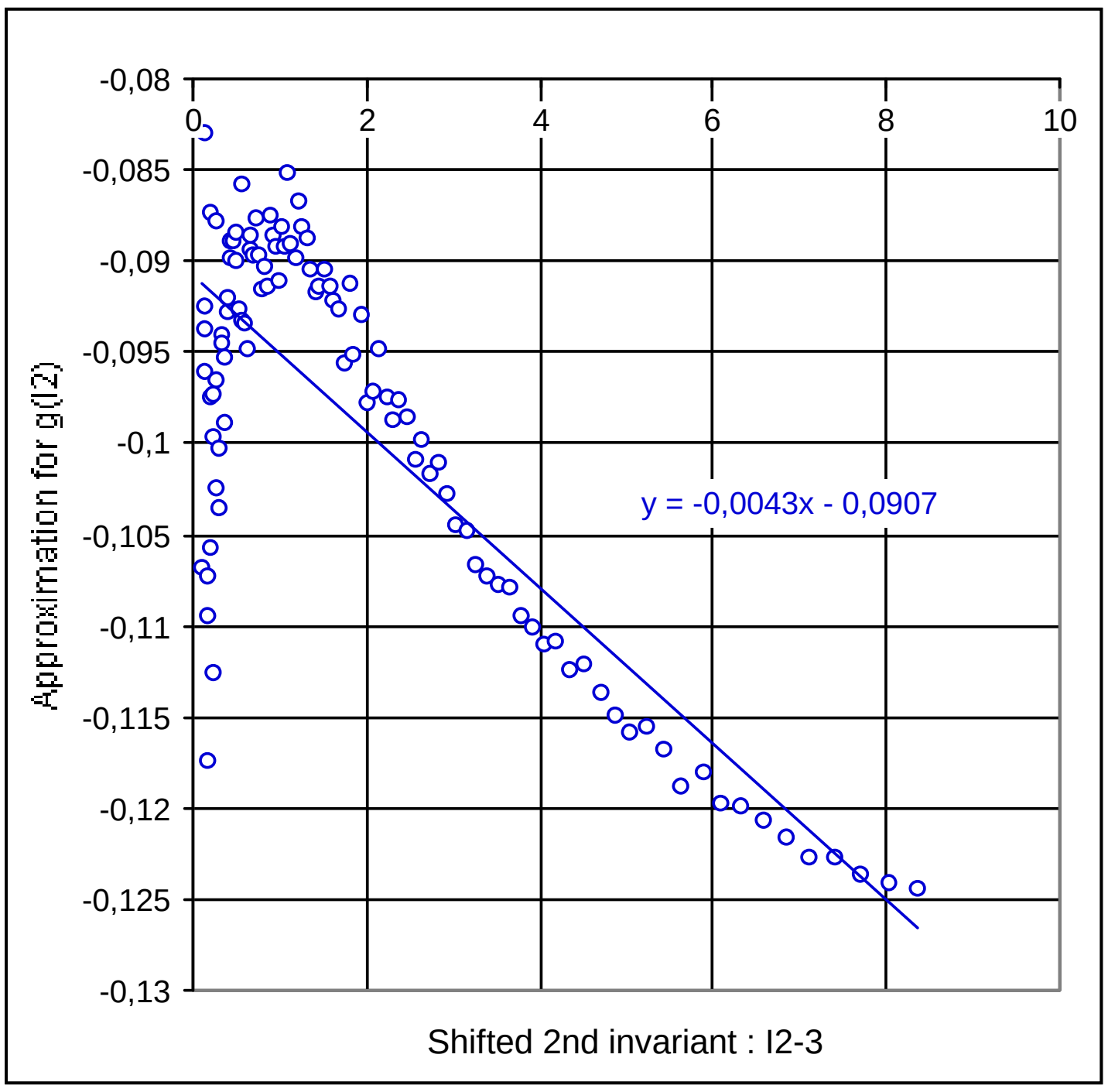

(a)

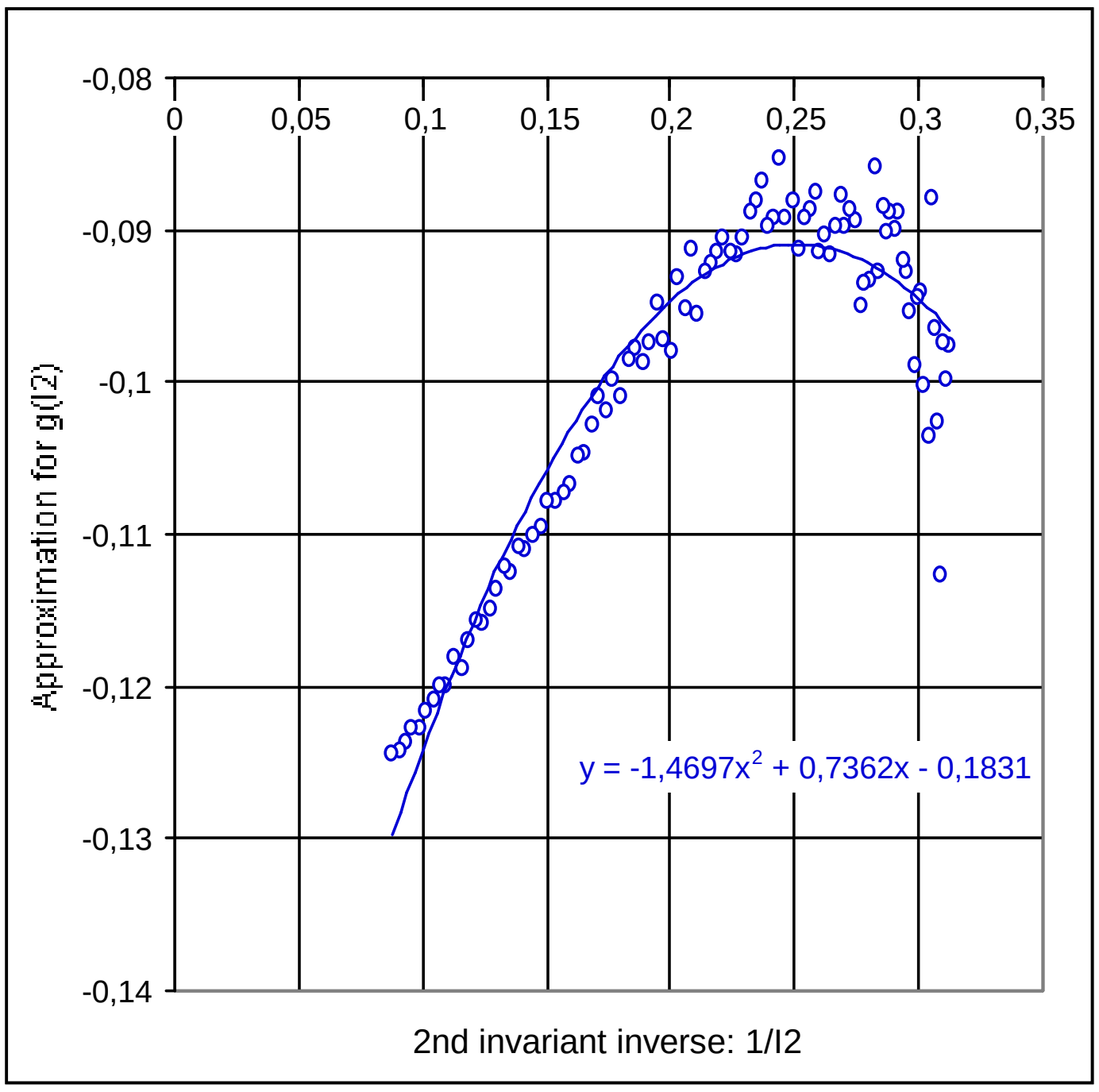

(b)

Figure 9: Second invariant identification.

(a) Hart-Smith form, (b) Rivlin form.

| | | | |
|---|---|---|---|
| $f(I_1)$ Rivlin form | $a_0$ = 0.274 | $a_1$ = -0.006 | $a_2$ = 0.0004 |
| $g(I_2)$ Hart-Smith form | $b_0$ = - 0.183 | $b_1$ = 0,732 | $b_2$ = -1.47 |

(a)

| | | | |
|---|---|---|---|
| $f(I_1)$ Rivlin form | $a_0$ = 0.20 | $a_1$ = -0.002 | $a_2$ = 0.0003 |
| $g(I_2)$ Hart-Smith form | $b_0$ = 0.075 | $b_1$ = 0.75 | $b_2$ = -1.47 |

(b)

Table 1: Material parameters for f and g functions before (a) and after (b) adjustment.

| Author(s) | Hyper-elastic potential | Material parameters |
|---|---|---|
| Mooney (1940) | $W = C_1(I_1-3) + C_2(I_2-3)$ | $C_1$ = 0.305 MPa<br>$C_2$ = 0.065 MPa |
| Ogden (1972) | $W = \mu_i (\lambda_1^{\alpha i} + \lambda_2^{\alpha i} + \lambda_3^{\alpha i})$ for i=1,2 | $\mu_1$ = 1.37 MPa<br>$\alpha_1$ = 1.18<br>$\mu_2$ = 0.003 MPa<br>$\alpha_2$ = 4.97 |

Table 2: Hyper-elastic potential expressions and identified values for Mooney and Ogden models.

| Model | $\Delta$ (%) | Number of material parameter |
|---|---|---|
| Mooney | 5 % | 2 |
| Ogden | 4 % | 4 |
| Our identified model | 0.1 % | 6 |

Table 3: Distance $\Delta$ between experimental data and fitted model expressions, case of Mooney, Ogden and our identified model.

**Adjustment**

Even tough the contribution of $g(I_2)$ is minimal during a tension test, it appears that experimental data and simulation don't fit perfectly. The last step of our procedure is to minimize the distance function defined previously (Eqn. 8). Since the initial estimation proposed with f and g functions is not very far from the experimental curve, the minimization procedure converges successfully. Table 1 summarizes numerical values of $a_i$ and $b_i$ material coefficients before and after the final minimization. We can see the perfect agreement. Both functions f and g are plotted on Fig. 10, which confirms that the second invariant contribution is small in tension where f increases very much as g decreases. Variations in function g occur in compression when function f remains quasi constant.

The values of table 1b are used to plot the global (compression and tension) behavior (continuous line) superposed with experimental data (circles in Fig.11). In order to have a full comparison, we also plot the Mooney and Ogden best fits on the same chart. The first model gives accurate description of the behavior in the compression range ($\lambda$<1). The second is better for describing the tension part of the behavior ($\lambda$>1). The potential expression and the material constants are given in table 2 for both models.

To quantify the quality of the adjustment, one can examine the distance $\Delta$ for each model in table 3. It appears that Mooney or Ogden gives the order of quality with 5% distance from experimental data. Our proposed choice for identification gives a more accurate representation ($\Delta$ = 0.1 %) even though 6 parameters are needed instead of 4 or even 2 for the Mooney form. It is worth noting that the same procedure used with only two parameters for $f(I_1)$ and 2 parameters for $g(I_2)$ leads to a $\Delta$ value of 1%. With only one parameter for each function the model becomes Mooney-like and leads to 5% distance.

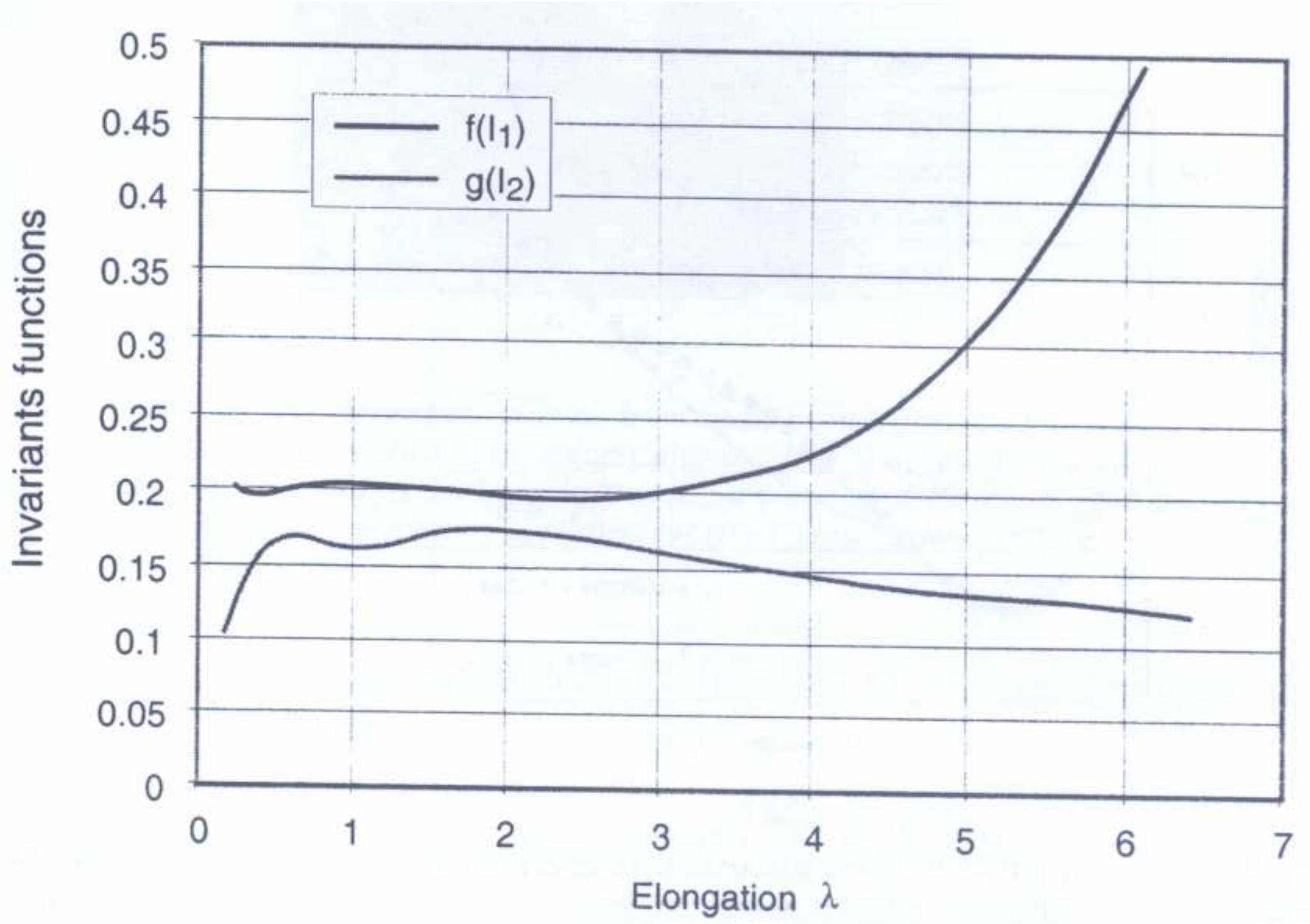

Figure 10: Comparison of the two invariant functions versus elongation $\lambda$. After adjustment it appears that $f(I_1)$ reaches higher values than $g(I_2)$ specially in uniaxial tension. The $g(I_2)$ function varies in compression but decreases slowly in tension so that the identification of $f(I_1)$ only can give an accurate representation of uniaxial tension but cannot predict compressive behavior.

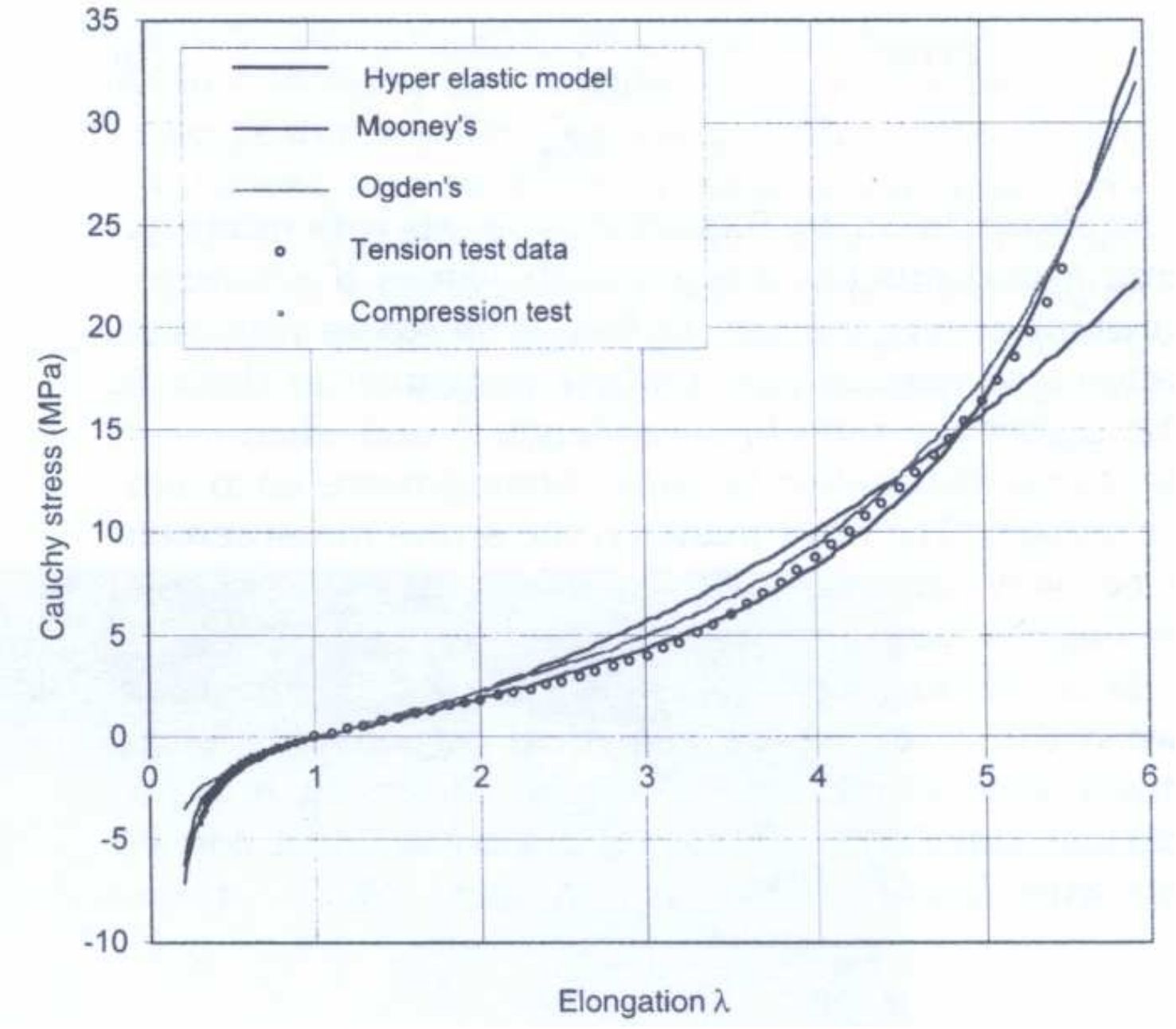

Figure 11: Tension and compression data compared with fitted models.Mooney's identification (2 parameters) failed to represent data in tension, the prediction is to high for small elongation and to low for high elongation. Ogden's identification (4 parameters) gives a good representation of tension data but failed to represent the compression behavior. Our model (6 coefficients) gives as good representation of tension or compression data.

## 3. *Complementary tests*

### *3.1 Plane strain test: a critical view of a classical test*

A common method to complete experimental data, when using a conventional uniaxial testing machine, consist in loading large specimen (initial width $L_o$) with respect to the initial length $h_o$ (see Fig. 12). Since the ratio h/L remains "small" a large area of the specimen is in a plane strain state (i.e., $\lambda_2 = 1$). This test is typically used with video control of the plane strain state by following dots in the middle part of the specimen. In that case, one doesn't follow edge effects ($\sigma_2 = 0$) that induce perturbations. These perturbations propagate inside the specimen over a length $\delta$ on both sides. This can be shown either numerically or experimentally.

#### Numerical simulation of plane strain test

Numerical simulations of the "plane strain" tests have been carried out using the Abaqus code. The hyper-elastic potential identified on the tension and compression tests has been implemented in a user subroutine UHYPER.f. Hybrid elements are used to control the incompressibility restriction. The initial geometry respects the small value of $h_O/L_O$ equal to 4/100. $h_O$ is stretched from 4 to 24 mm. A typical result of the simulation (Fig.13) shows the validity of the "plane strain" in the central region ($\varepsilon_2$ nominal strain component is equal to '0'.) The Cauchy stress component $\sigma_1$ is perturbed with edge effects on a length $\delta$ that is of the order of the specimen length h. In this zone the axial Cauchy stress decreases from $\sigma_1$ to approximately $\kappa\sigma_1$ (with $\kappa$ value equal to 0.6 at the beginning of the step).

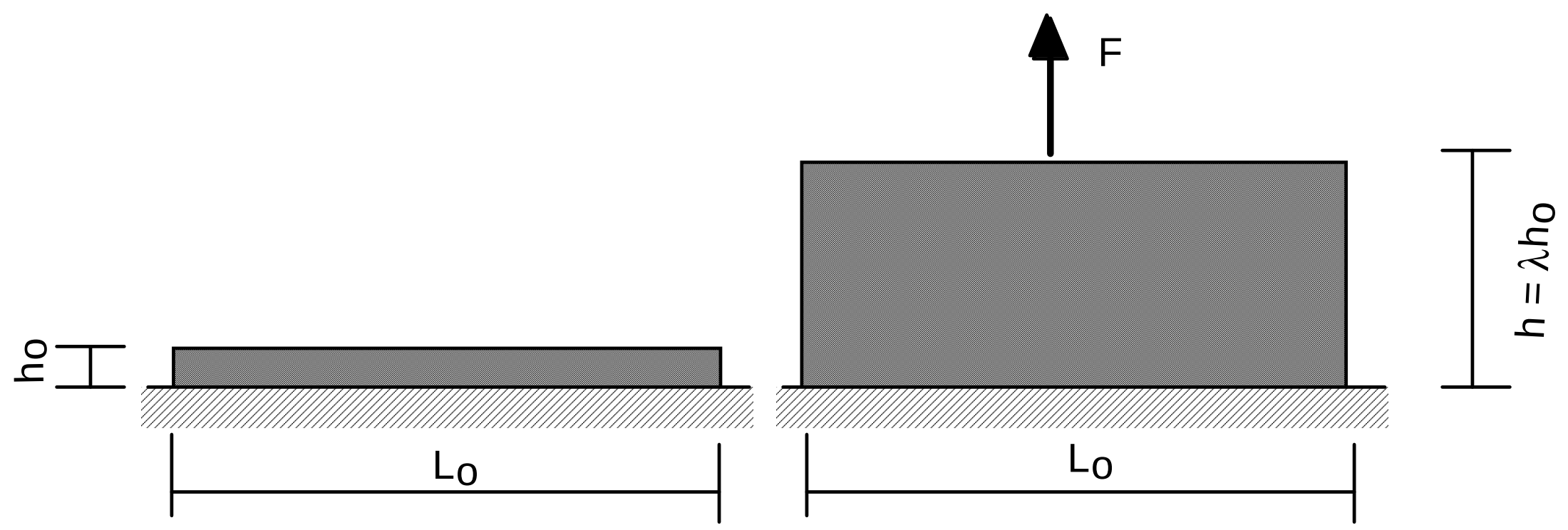

“plane strain” tension : $h_0/L_0$ as small as possible

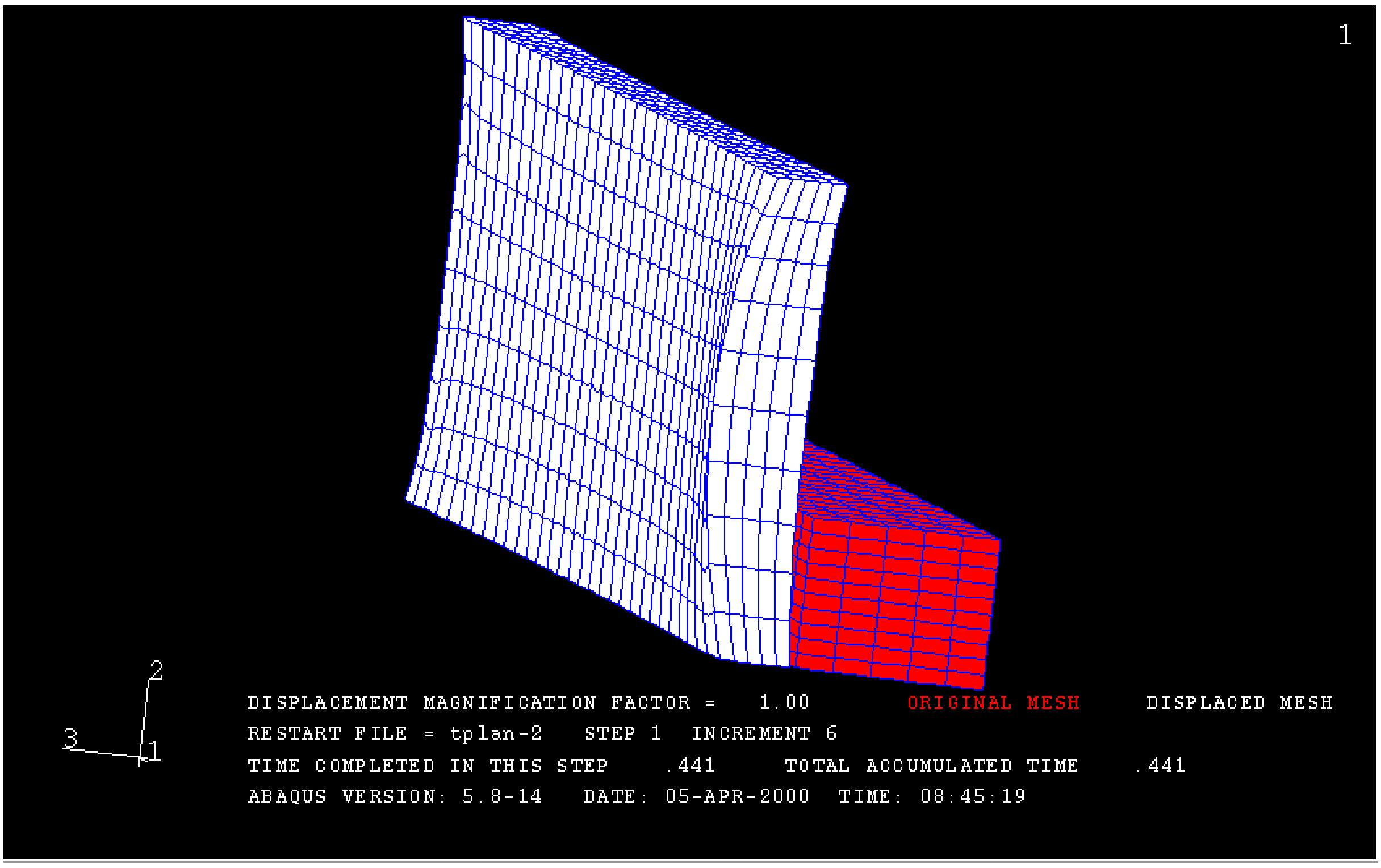

Figure 12: « Plane strain » test, principle and ABAQUS simulation ($\lambda$ = 3)

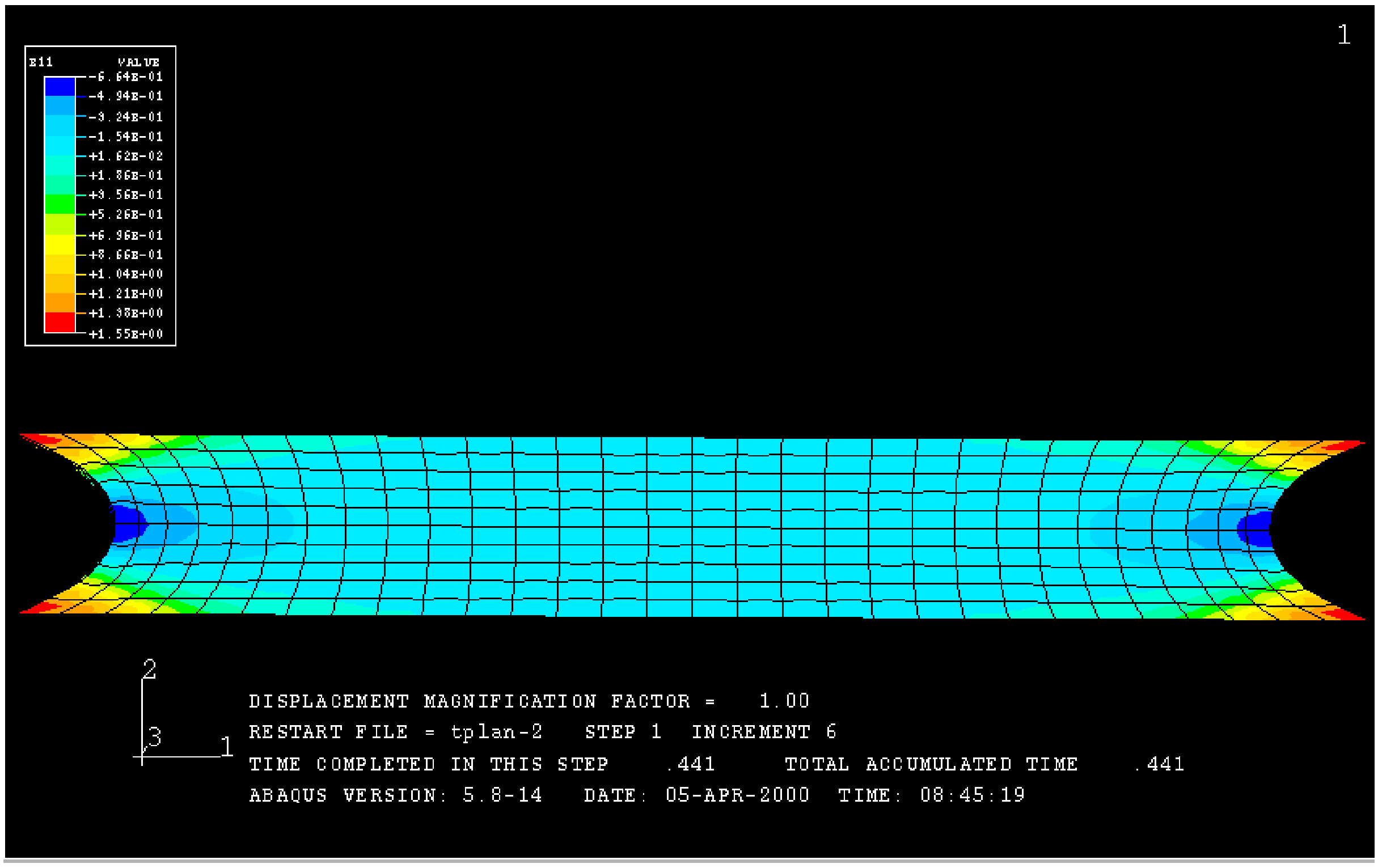

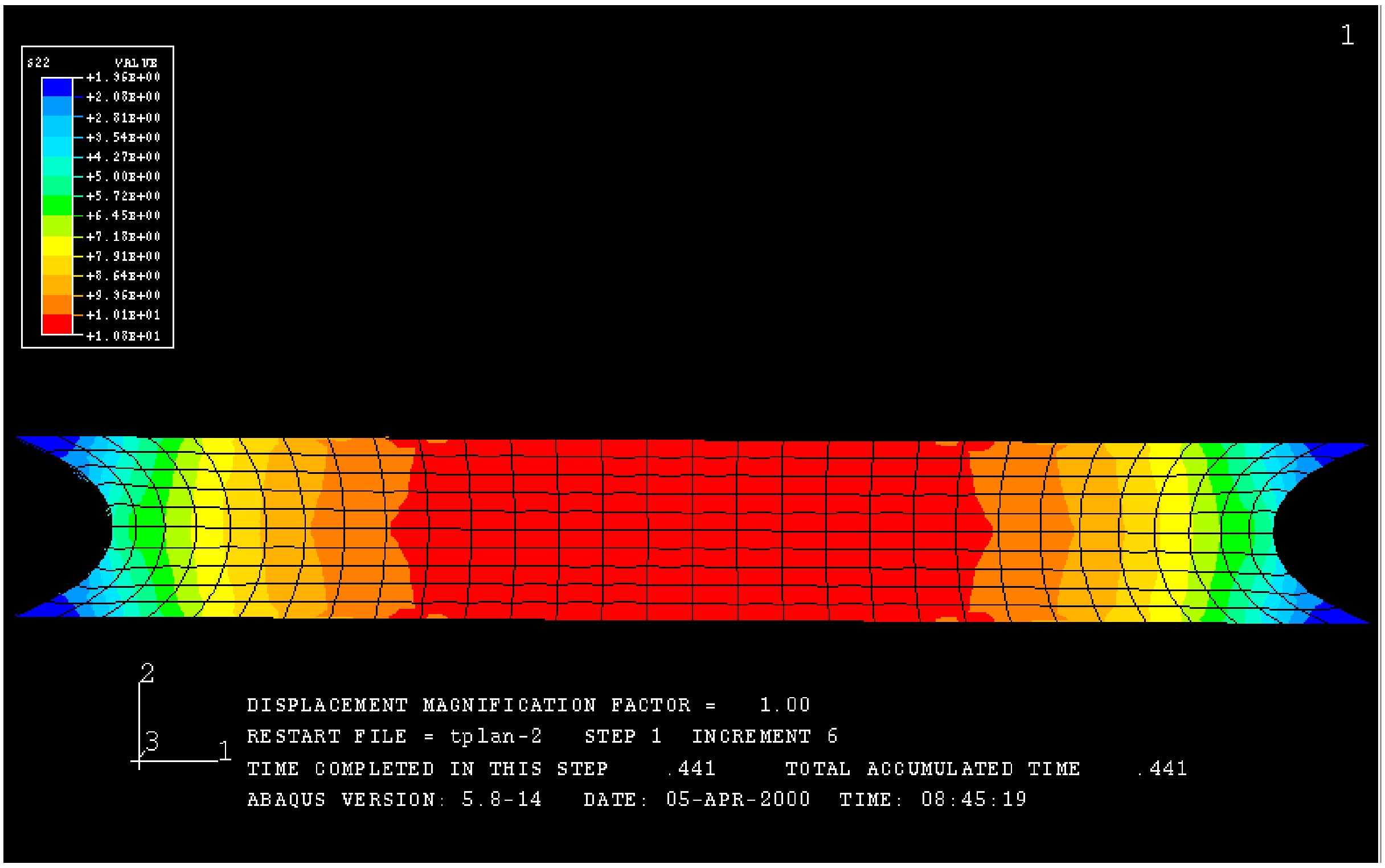

Figure 13: Numerical simulation of “plane strain” test with ABAQUS

(top) $\varepsilon_{22}$ strain distribution, (bottom) $\sigma_{11}$ stress distribution

#### Error evaluation

We define the stress error as the real value of $\sigma_1$ in the medium region of the specimen, compared with the measured one $\sigma_m$ which is the mean value of $\sigma_1$ over the all specimen.

$$Error = \frac{\sigma_1 - \sigma_m}{\sigma_m} \tag{15}$$

If we assume that tension stress vary from $\kappa\sigma_1$ to $\sigma_1$ over a length $\delta$, as it is illustrated in Fig. 14-a, the mean value $\sigma_m$ can be calculated. Then the error is deduced from Eqn. 15 and given in the following expression:

$$Error = \frac{1}{1+\left(\frac{2\lambda h_o(\kappa-1)}{3L_o}\right)} - 1 \tag{16}$$

As shown by Eqn. 16, this error increases with extension ratio $\lambda$ and quickly reaches important values if $\kappa$ remains constant during the test. In fact, $\kappa$ increases with elongation and reaches $\kappa$=1 when elongation is about 5. The higher the ratio h/L, the larger $\delta$ and when $\lambda$=5 the stress distribution becomes homogeneous as in simple tension. The error made on the stress measurement is not very important (Fig. 14-b) but the strain state is no longer a "plane strain" configuration.

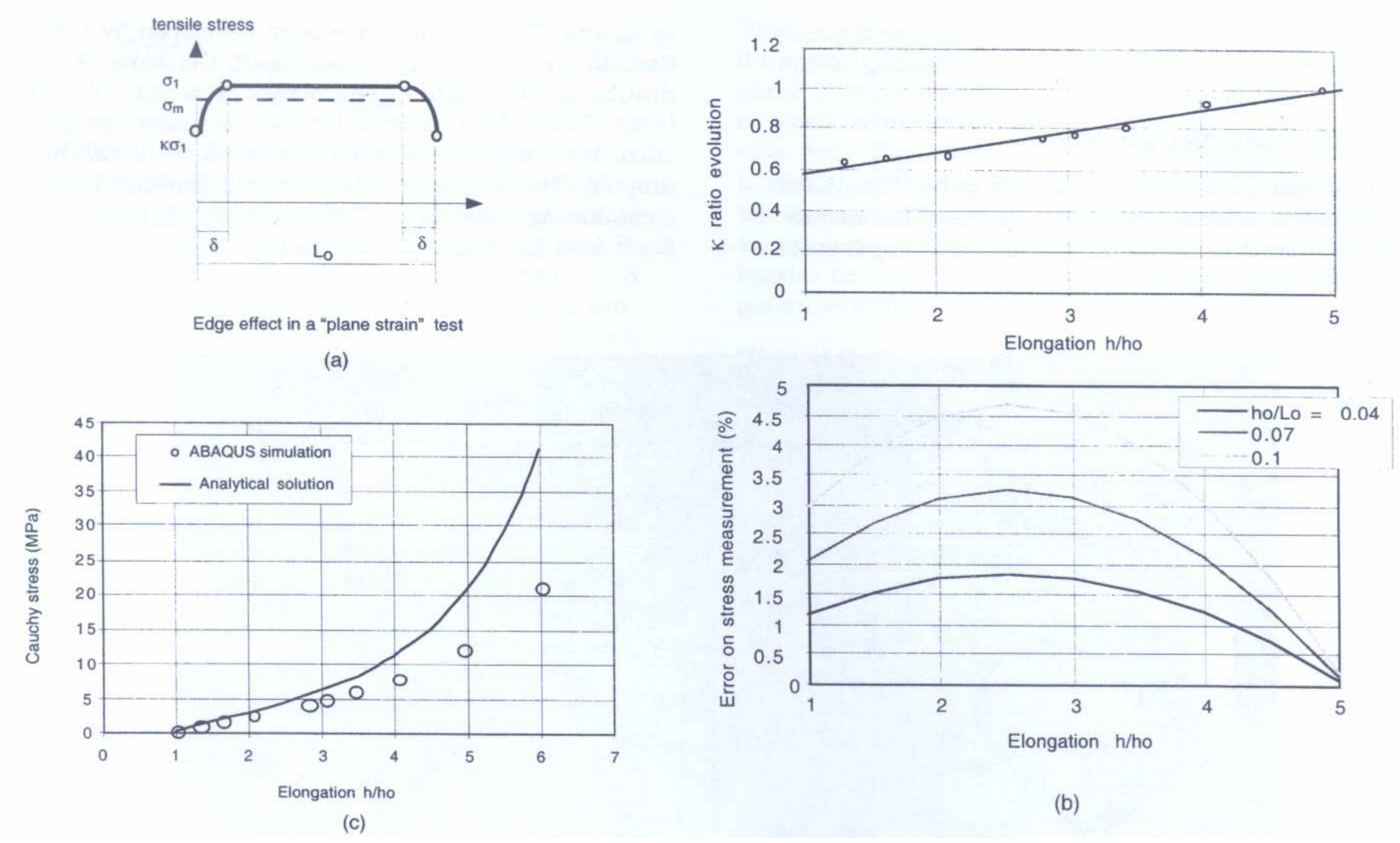

Figure 14: Error on mean stress value during a 'plane strain' test

As an illustration Fig. 14-c shows the difference between the results of the analytical response for plane strain state (continuous line) and the result of a numerical simulation of the "plane strain" test made with Abaqus. The circles represent the stress-strain response in the central part of the specimen. It appears that numerical simulation begins like a plane strain response but quickly diverges to lower stresses values more representative of a simple tension state. This kind of test should not be used for potential identification. Besides, invariant $I_1$ and $I_2$ are equal so that the effect of function g is not dominant for such a test. Consequently, it is chosen to develop a biaxial test validation procedure of the hyper-elastic potential form.

### *3.2 Biaxial tension tests*

Several authors, for example Kawabata (15), Meissner (16) or Reuge (17) have developed biaxial apparatus to test rubber-like material in equi-biaxial tension. Our aim is to use simple biaxial tension on a specific multiaxial testing machine to validate the identified model. The possibilities of this equipment in term of biaxiality can be compared with the Meissner apparatus (biaxiality ratio can vary) but positioning of the specimen is much easier.

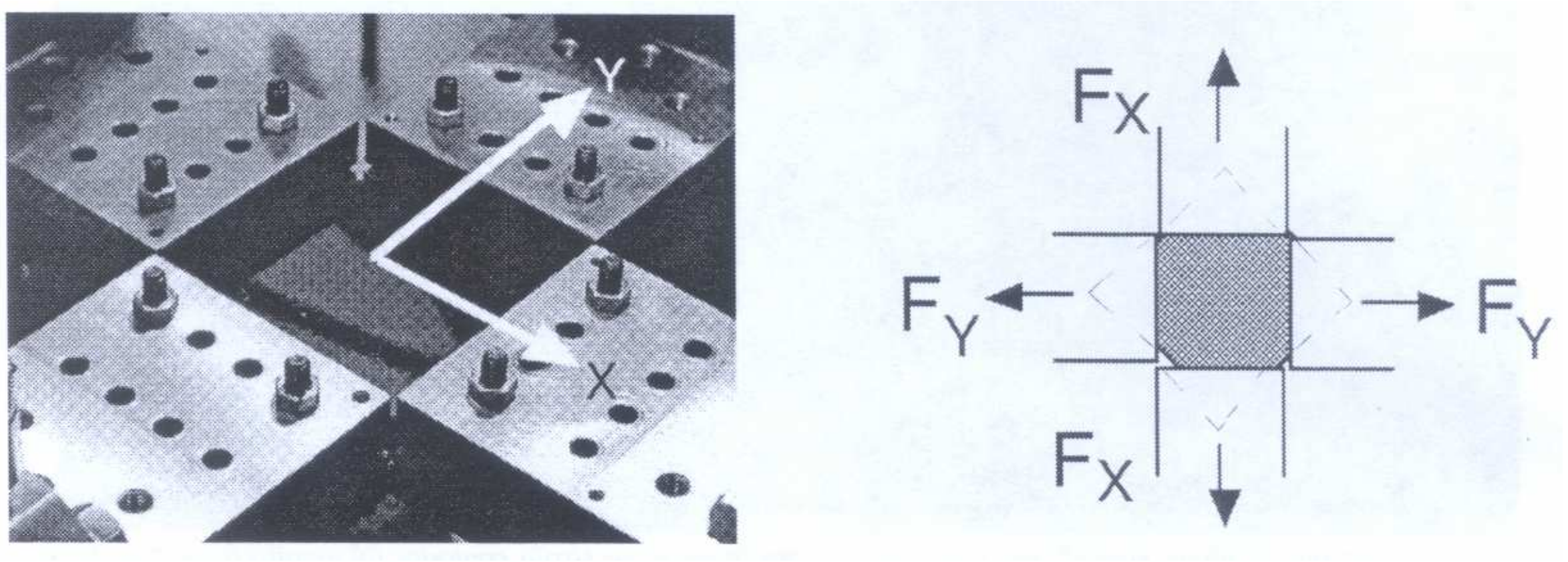

Fig 15: Biaxial testing on rubber-like material in the multiaxial machine ASTREE.

#### **Multiaxial testing machine Astree**

The biaxial tests presented in this paper have been carried out on a triaxial testing machine. This electro-hydraulic testing machine has six servohydraulic actuators. This machine has been developed by LMT-Cachan and Schenck AG, Darmstadt, Germany. An hydraulic power station generates a rate of flow of 330 l/mn. Closed loop control for each actuator is provided by a digital controller, Schenck 59 serial hydropuls. The controller monitors provides signal conditioning for each load actuator position channel. Each axis (X, Y and Z) has its own dedicated strain channel for signal conditioning and control. Strain input signals can come from a variety of strain measuring devices (for example, strain gauges, extensometer, etc.). Computer test control and data acquisition are accomplished by an object-oriented programming software (LabVIEW®).

Two tests have been carried out at room temperature (i.e., T=25°C) on square specimen. Fig. 15 shows how the square specimen is positioned in the testing area: an initial square is rotated with a 45° angle wrt. loading directions. A smaller square region is then isolated and can be stretched in different ways. In a first equi-biaxial tension test, the square specimen is simultaneously stretched in both directions X and Y. In a second test, we first stretch the specimen in the X direction and then in the Y direction in order to reach the same state of deformation. The results are discussed in the following section.

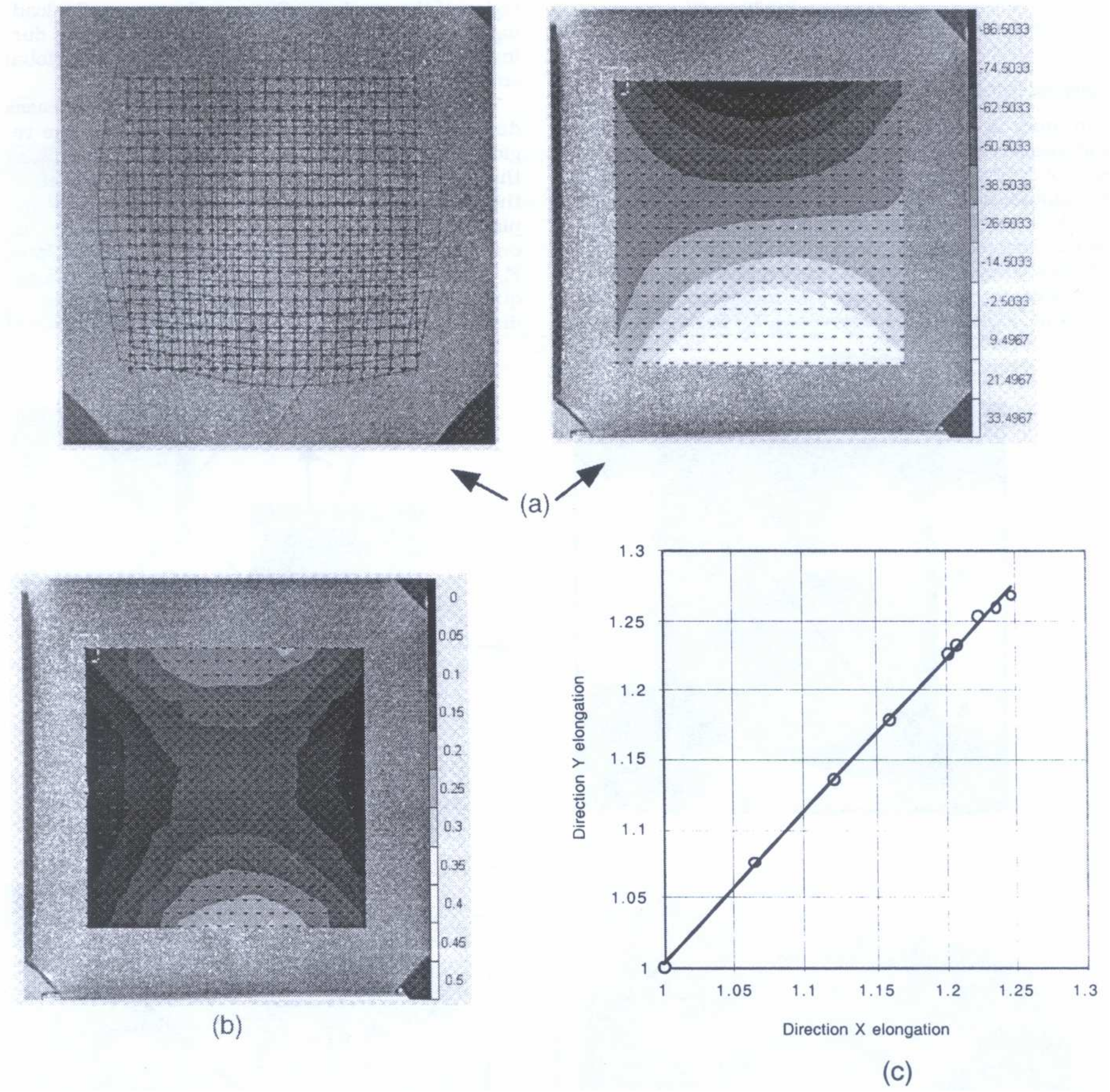

Fig 16: Biaxial stress measurement, displacement and strain field in a biaxial test

#### A quasi homogeneous test: equilibrate biaxial tension

Figure 16-a shows the displacement field after bi-axial testing on our rubber specimen. This field is not homogeneous, so that it is not possible to estimate local strain from the grip displacements. Figure 16-b shows the distribution of the principal elongation in the central region of the specimen. One can see that both $\lambda_x$ and $\lambda_y$ elongation are equal in that area and so this zone is in an equi-biaxial elongation state. We observe that the strain is quasi constant along AB (Fig. 16-c). If we assume that the stress components are quasi uniform in this section AB, we can estimate (Fig. 17) the Cauchy stress from the biaxial loading force F:

$$\sigma_{BT} = \frac{\sqrt{2}F}{e.L} \qquad (26)$$

'e' is the current specimen thickness and 'L' the current length between A and B. Figure 17-b shows that loading forces are equal in both directions. In Fig. 17-c the estimated stress from the loading force (circles) and the prediction (continuous line) of the previously identified hyper-elastic model in biaxial tension are superposed. This first approximation confirms the form of the second invariant function $g(I_2)$ which is predominant both in bi-axial tension as in compression.

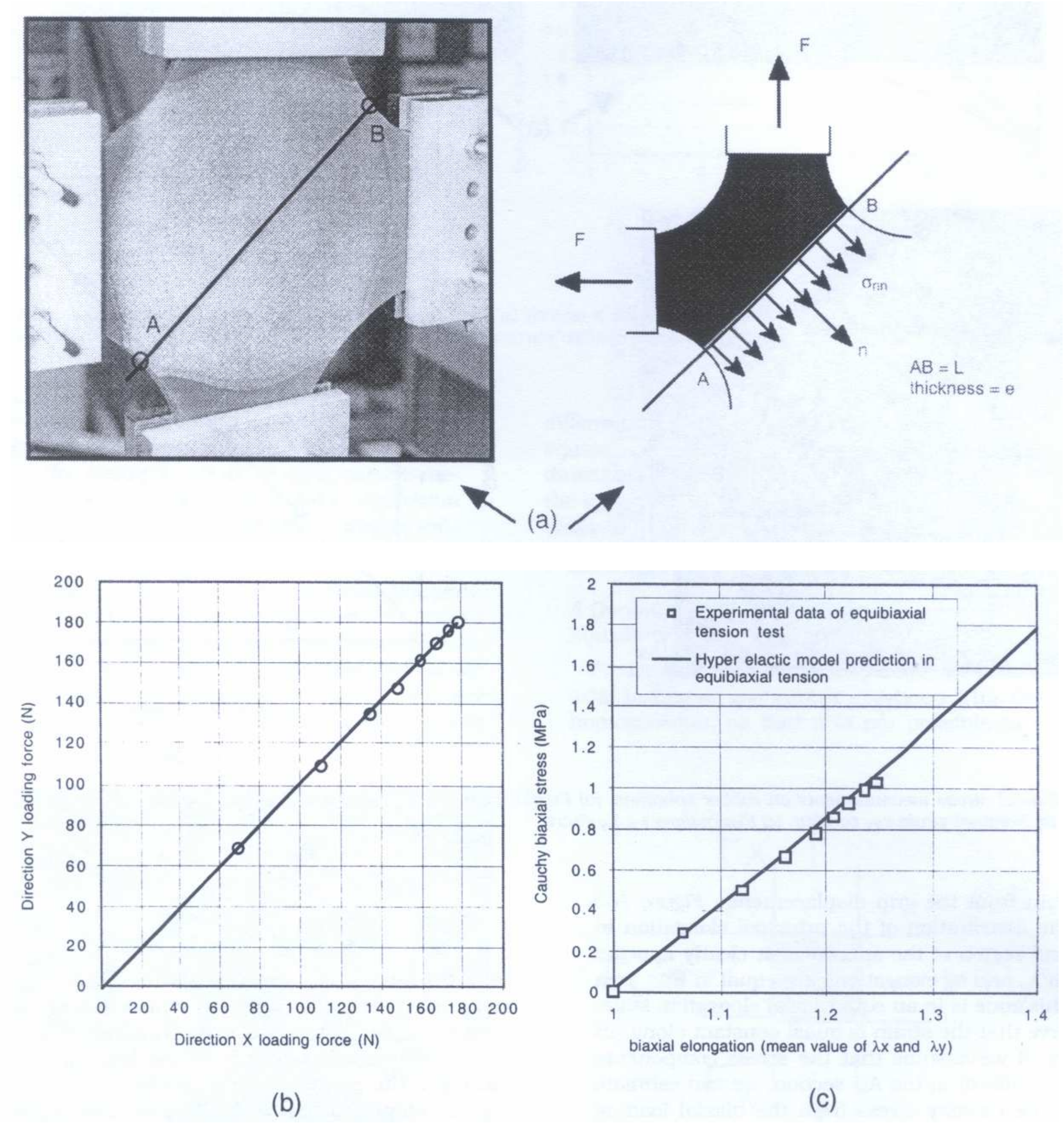

Fig 17: analytical identification of the equibiaxial behavior

**Complex path of deformation**

In order to make comparisons between simulated and measured displacement fields we apply a specific path of deformation. The first set of grips (direction X) is maintained fixe while the second set (direction Y) stretches the specimen. We then stop elongation in direction 2 and apply stretching in direction X. Figure 18 gives a representation of this loading path in the displacement chart ($U_x$, $U_y$), the measured loading forces ($F_x$, $F_y$) and the elongation ($\lambda_x$, $\lambda_y$) in the central region of the specimen. One can observe on the loading force chart that $F_y$ remains quasi constant during the second step, this can be surprising since global strain increases.

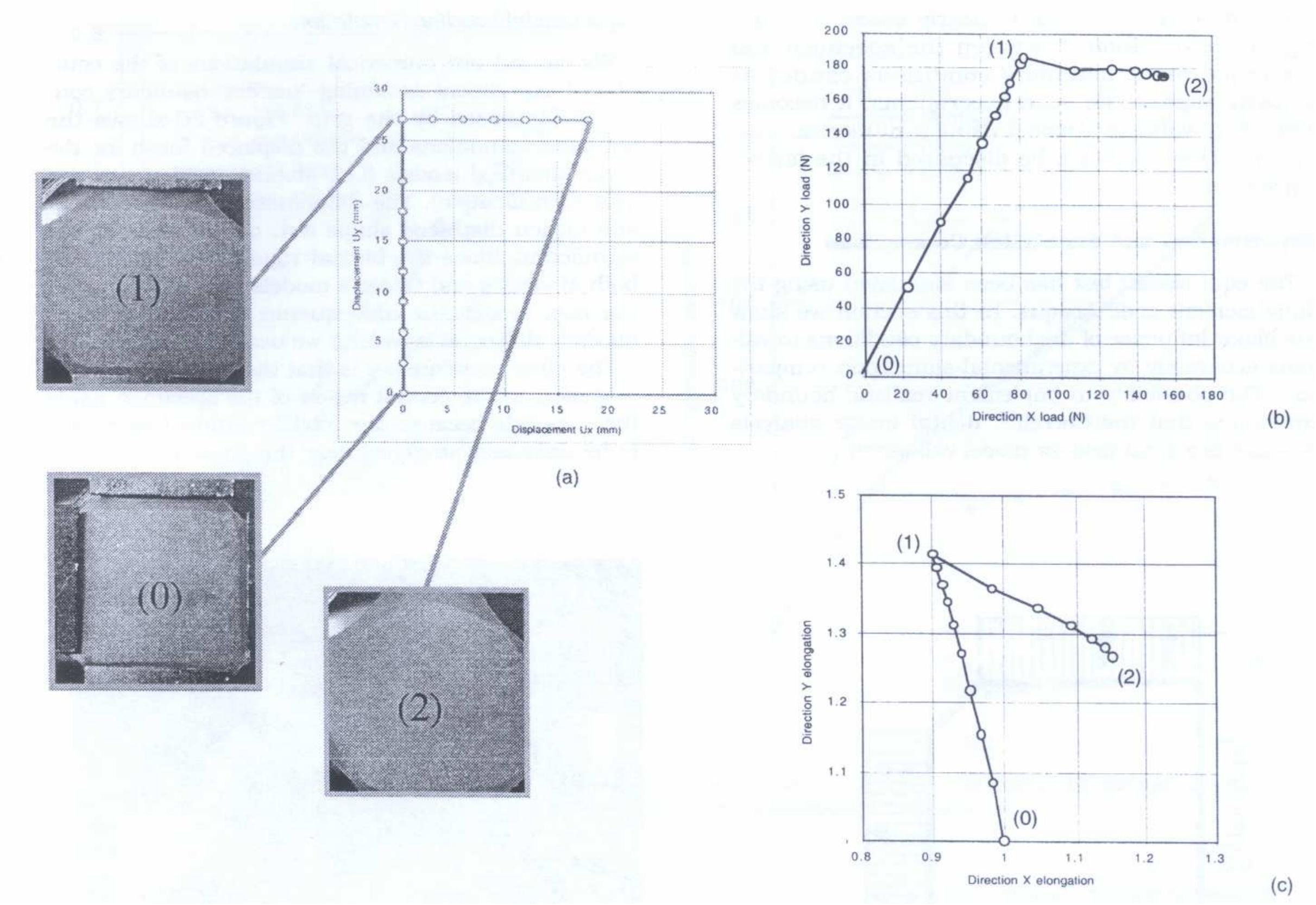

Fig 18: Sequential biaxial test

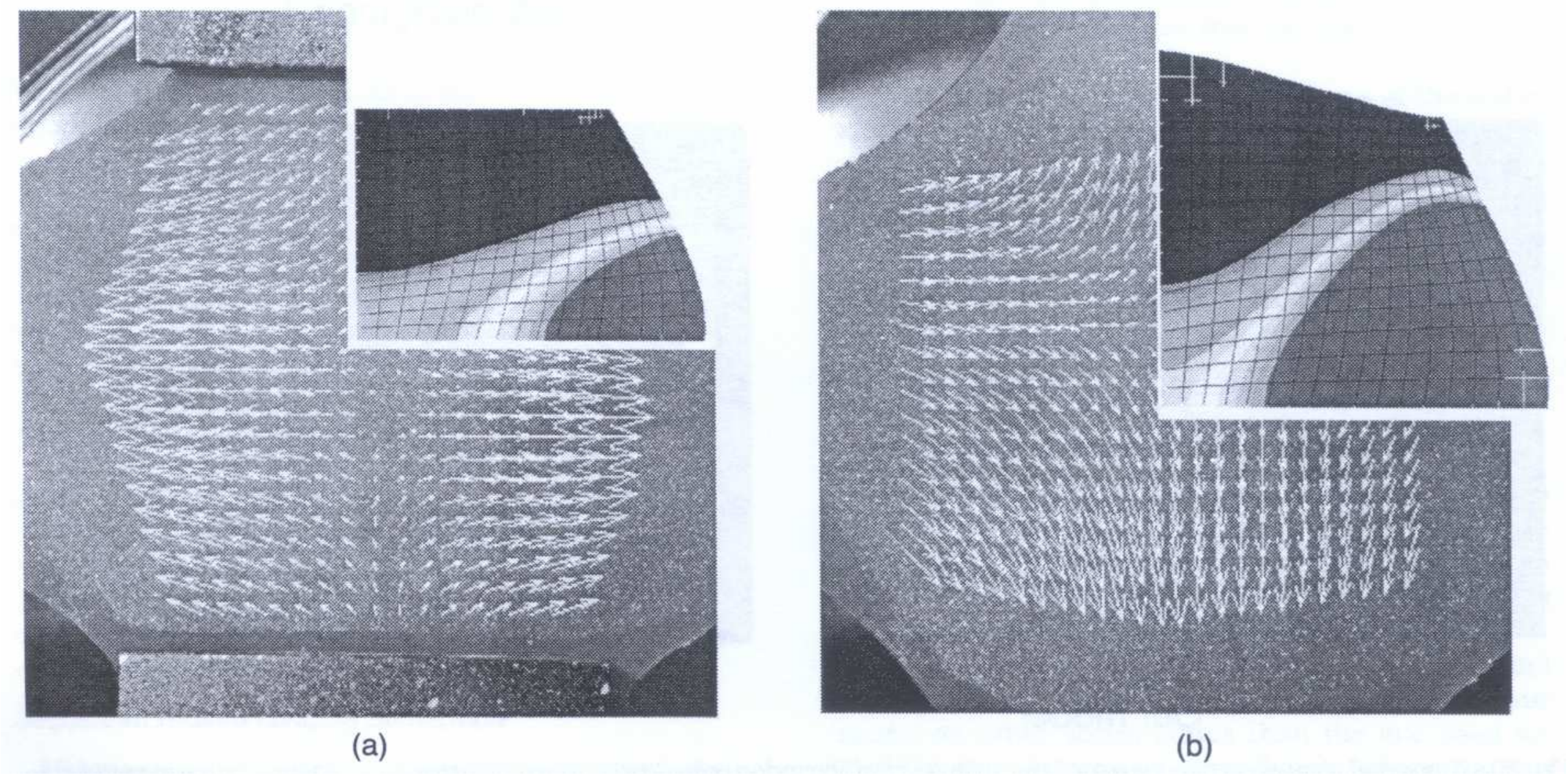

Fig 19: Displacement field at the end of step 1 and step 2

The elongation $\lambda_x$ in the central region decreases during the first step. This can mean that the regions near the grips are elongated in direction X, or this can be due to the sliding of the specimen in the grip. Sliding is confirmed by the displacement field shown on figure 19-a. During the second stage, sliding continues and the loading force $F_y$ does not increase as we could expect in a sequential biaxial loading. Figure 19-b represents the displacement field at the end of the second stage wrt. end of the first step: it clearly shows the sliding in the direction Y between the specimen and the grips. Since boundary conditions are not well imposed for such experiments, it becomes difficult to validate a model using a numerical simulation. This point will be discussed in the following section.

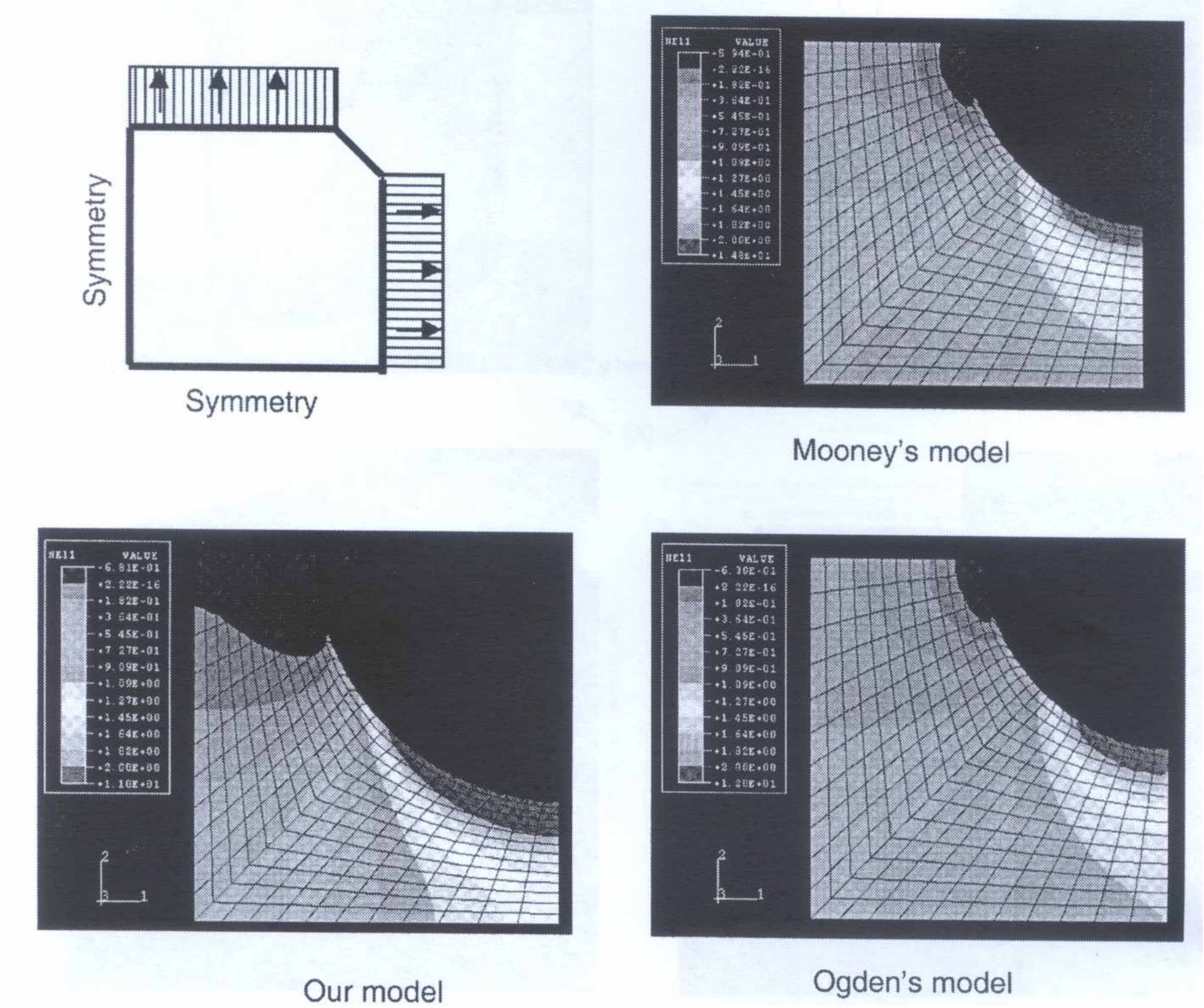

Fig 20: Numerical simulation of equibiaxial test with 'perfect' boundary conditions
Experimental and simulation comparison

The equi-biaxial test has been simulated using the finite element code Abaqus. In this section we show the major influence of the boundary conditions to validate accurately by experimental-simulation comparison. The possibility, to implement realistic boundary condition that the Correli$^{GD}$digital image analysis provides is a great help for model validation.

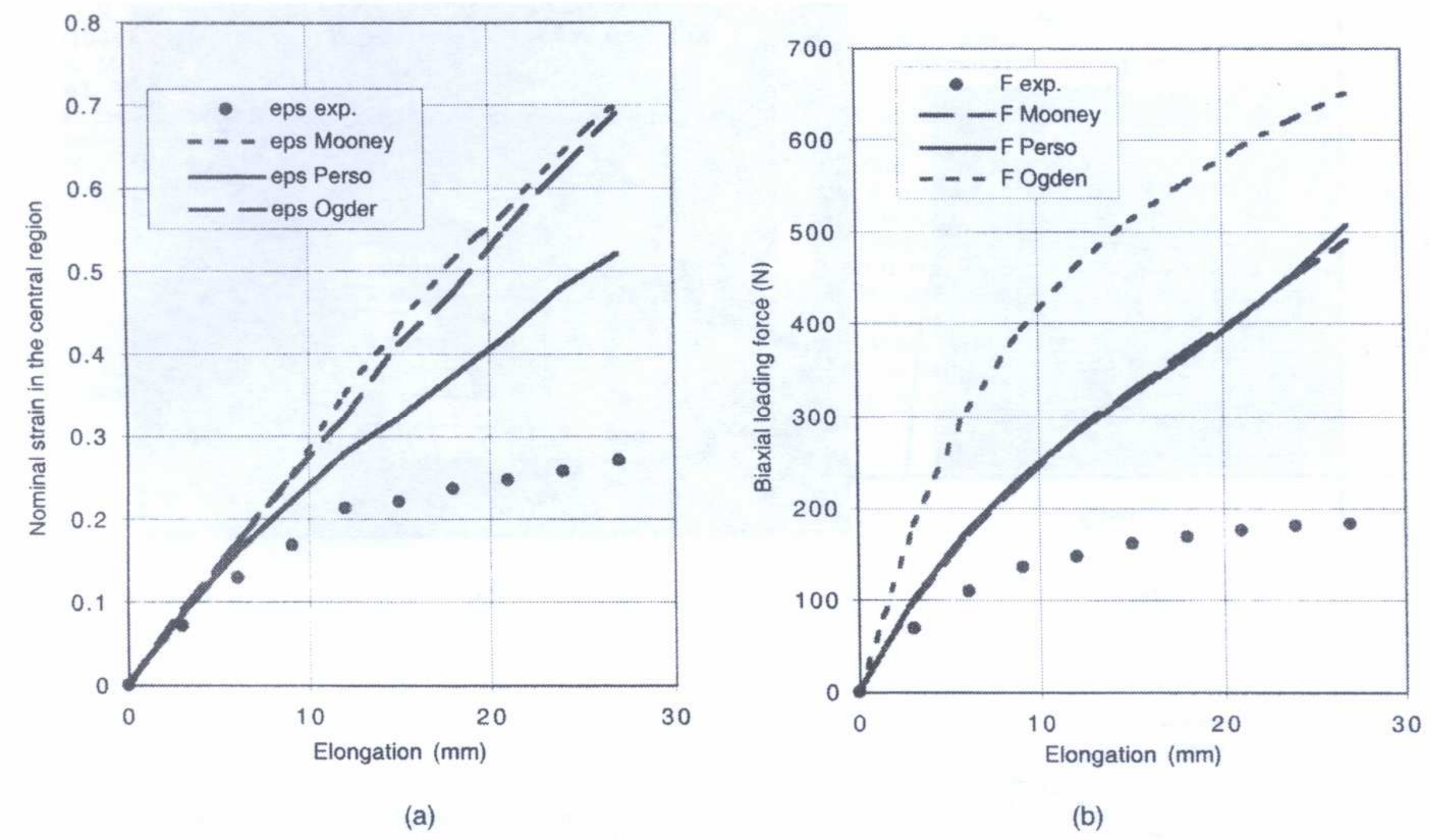

Fig 21: Numerical simulation results compared with experimental data for ‘perfect’ boundary conditions.

**Equi-biaxial loading simulation**

We carried out numerical simulation of the equi-biaxial experiment assuming 'perfect' boundary conditions imposed by the grip. Figure 20 shows boundary conditions and the displaced mesh for the 3 identified models (i.e., Mooney, Ogden and our identification). The differences between Mooney and Ogden displaced shape wrt. our model are significant. Since the biaxial rigidity predicted with both Mooney and Ogden is lower than by our model, the global consequence is that the region biaxially deformed is wider if we use such models.

The other consequence is that the biaxial elongation is greater in the central region of the specimen using these models because our model demands less energy to be uniaxially stretched near the grips. This, in itself could be a determining criterion to discriminate between models. However, the experimental data are very far from the simulated force (Fig. 21-a) or central elongation (Fig. 21-b). If we compare the simulated displacement field with the measured one, it appears that the 'perfect' boundary conditions are not realistic.

If we impose boundary conditions directly issued from the inter-correlation analysis, as illustrated on Fig. 22-a, the differences in terms of elongation are smaller (less than 1%). This is easy to understand since the real displacement field is prescribed on the edge and incompressibility condition is maintained. Moreover, Fig. 22-b shows the stress distribution which valid the assumption of quasi uniform stresses along AB. This confirms a posteriori the good agreement at local scale between the analytical biaxial behavior and the experimental measures in the central region of the specimen.

Finally, Fig.22-c shows the biaxial force versus elongation in the central region plot for both numerical simulation and experimental data. The global error between these results is less than 5% which is quite good for such simple experiment. This error is of the order of the error made between elongation $\lambda_x$ and $\lambda_y$ values so we cannot be more precise.

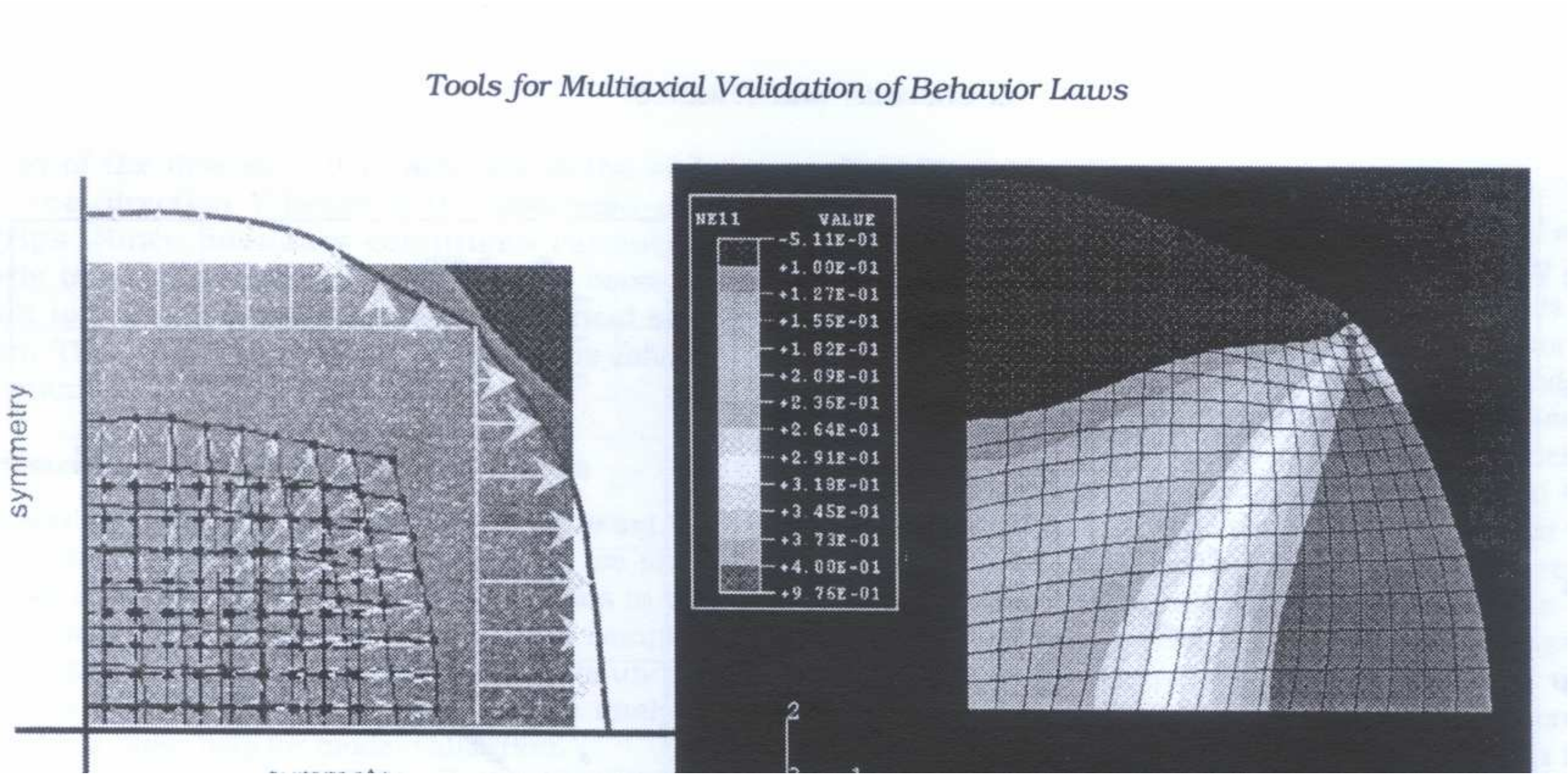

Fig 22: Numerical simulation results compared with experimental data for boundary conditions issued of Correli$^{GD}$

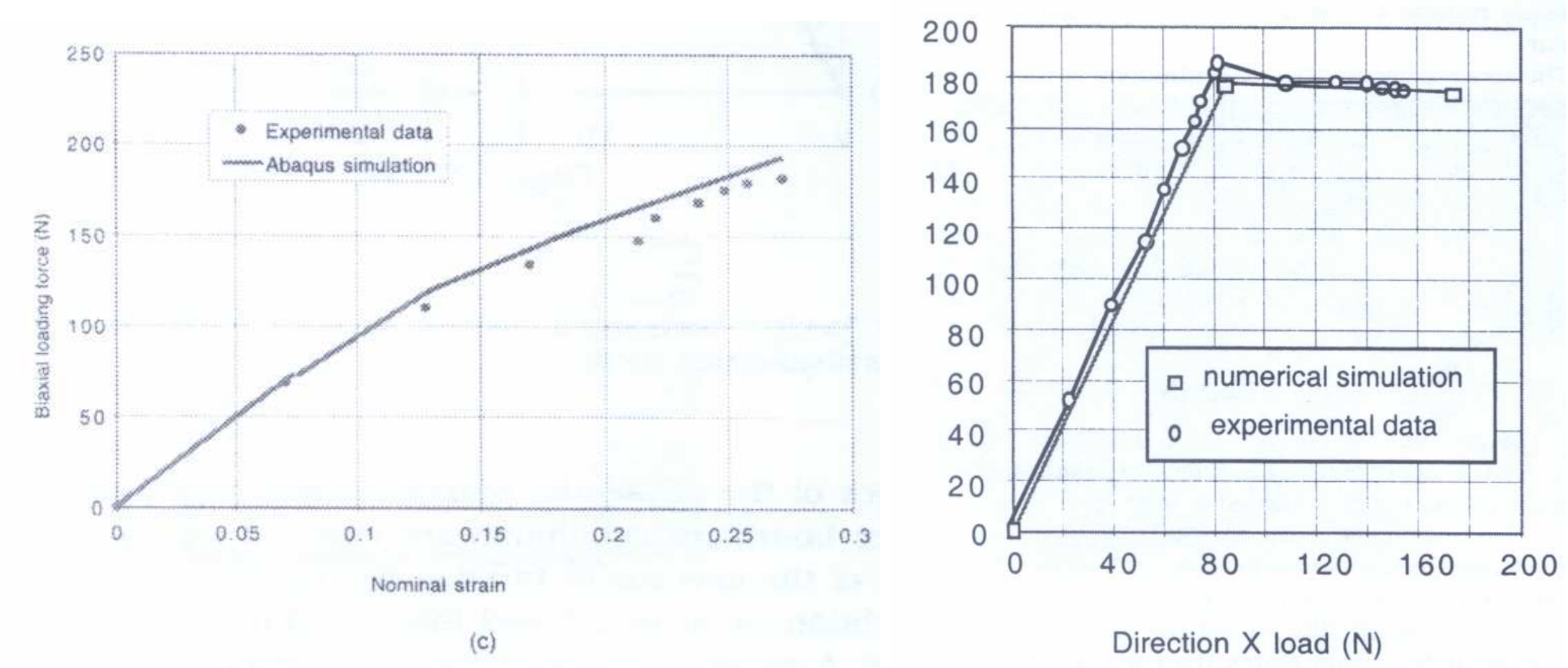

Fig 23: Numerical simulation compared with experimental data for sequential biaxial test.

#### Sequential biaxial loading simulation

Using realistic boundary conditions issued from the intercorrelation analysis allows simulation of both steps of the sequential biaxial-loading experiment. Since boundary conditions are imposed on a large part of the specimen's frontier, the incompressibility condition must be relaxed to avoid numerical artifacts. A slight compressibility is implemented in the Abaqus routine to make sure no such instabilities can occur.

The strain component $\varepsilon_{11}$ distribution at the end of step 1 and step 2 are superimposed onto the experimental deformed mesh of Fig. 19-a and 19-b. Finally, Fig. 23 shows good matching between the biaxial elongation in the central region at the end of both steps issued from numerical simulation compared with experimental results. It also shows good matching in terms of loading force prediction for this complex path.

## *4. Conclusions*

A specific identification procedure has been used to identify the hyper-eleastic behavior of rubber-like materials. Tension and compression tests have been performed and validated using a digital image analysis to follow large displacements.

Biaxial tests have been carried out on a multiaxial testing machine ASTREE to test the prediction of our model on other strain states than the one used for identification. Due to strain heterogeneity the validation must be done by confrontation with numerical simulations. Our identified model has been implemented in Abaqus and, when realistic boundary conditions are specified, gives accurate prediction of both equi-biaxial and sequential biaxial tension tests.

Future work in this field will deal with the development of specific grips to reduce sliding and allow larger biaxial strains. Model discrimination would ten be easier for larger values of the biaxial elongation. Loading and unloading effects will be studied as well as in uniaxial and biaxial loadings. The study of the damage that occurs when the material reaches very large strain will be our next subject of interest in this.

## *Acknowledgment*

SMAC-Toulon provided us with Smactane™ sheets. Special thanks for Laura and Pierre for English editing